\documentclass[11pt,a4paper,twoside]{paper}

\usepackage[T1]{fontenc}
\usepackage[latin1]{inputenc}
\usepackage{amsmath,amssymb,amsthm}
\usepackage{graphicx}
\usepackage{parskip}
\usepackage[margin=1in]{geometry}
\paragraphfont{\sffamily\itshape}

\newtheoremstyle{etoile}{\parskip}{\parskip}{\itshape}
                        {0pt}{\bfseries\sffamily}{.}{ }{}
\theoremstyle{etoile}

\newtheorem{prop}{Proposition}[section] 

\newtheorem{theo}[prop]{Theorem}

\newtheorem{lem}[prop]{Lemma}

\makeatletter
\@addtoreset{equation}{section}
\makeatother

\newcommand\egaldef{\stackrel{\mbox{\upshape\tiny def}}{=}}

\newcommand\1{\leavevmode\hbox{\rm 
\small1\kern-0.35em\normalsize1}}
        
\newcommand\EE{\mathsf{E}}
 
\newcommand\Gb{\mathbf{G}}

\newcommand\Oc{\mathcal{O}}

\newcommand\ZZ{\mathbb{Z}}
\newcommand\n{^{\scriptscriptstyle (N)}}

\newcommand\nm{^{\scriptscriptstyle (N-1)}}
\newcommand\nmm{^{\scriptscriptstyle (N-2)}}
\def\DD{\displaystyle} 
\DeclareMathOperator*{\lra}{\leftrightarrows}
\DeclareMathOperator*{\rla}{\rightleftarrows}
\DeclareMathOperator*{\ra}{\rightarrow}

\begin{document}
\title{Stochastic Dynamics of Discrete Curves and Multi-Type
Exclusion Processes}

\author{Guy Fayolle \thanks{INRIA - Domaine de Voluceau, 
Rocquencourt
BP 105 - 78153 Le Chesnay Cedex - France. Contact:
\texttt{Guy.Fayolle@inria.fr, Cyril.Furtlehner@inria.fr}} 
\and Cyril
Furtlehner \footnotemark[1]} 
\date{January 2006}

\maketitle
\abstract{This study deals with continuous limits of interacting one-dimensional diffusive systems, arising from stochastic distortions of discrete curves with various kinds of coding representations. These systems are essentially of a reaction-diffusion nature.  In the non-reversible case, the invariant measure has generally a non Gibbs form. The corresponding steady-state regime is analyzed in detail with the help of a tagged particle and a state-graph cycle  expansion of  the probability currents. As a consequence, the constants  appearing  in Lotka-Volterraequations ---which describe the fluid limits of stationary states--- can be traced back directly at the discrete level to tagged particles cycles coefficients. Current fluctuations are also studied and the Lagrangian is obtained by an iterative scheme. The related Hamilton-Jacobi equation, which leads to the large deviation functional, is analyzed and solved in the reversible case for the sake of checking.}

\newpage
\tableofcontents
\newpage
\section{Introduction}\label{INTRO}
Interplay between discrete and continuous description is a recurrent
question in statistical physics, which in some cases can be addressed quite rigorously via probabilistic methods. In the context of reaction-diffusion systems this amounts to studying fluid or hydrodynamic limits, and number of approaches have been proposed, inparticular in the framework of exclusion processes, see
\cite{Li},\cite{MaPr} \cite{Sp}, \cite{KiLa} and references therein.
As far as the above limits are at stake, all these methods have in
common to be limited to systems having stationary states given in
closed product form, or at least to systems for which the invariant
measure for finite $N$ is explicitly known. For instance,
\textsc{asep} with open boundary are described in terms of matrix
product forms (really a sort of non-commutative product form), and the
continuous limits can be understood by means of Brownian bridges
\cite{DeEnLe}. We propose to tackle these problems from a different
view-point. The initial objects are discrete sample paths enduring
stochastic deformations, and our primary concern is to understand the
nature of the limit curves, when $N$ goes to infinity: how do they
evolve in time, and which limiting process do they represent as $t$
goes to infinity: in other words, what are the equilibrium curves?
Following \cite{FaFu} and \cite{FaFu2}, we give here some partial
answers to these questions.

In \cite{FaFu} a specific model was considered, namely  paths on the square 
lattice, and we could reformulate the problem in terms of coupled exclusion 
processes, to understand the thermodynamic equilibrium and a phase 
transition point above which curves reach a deterministic profile, solution 
of a nonlinear dynamical system which was solved explicitly by means of 
elliptic functions. Two extensions of this system
were introduced in  \cite{FaFu2} :
\begin{itemize}
\item one which comprises multi-type exclusion particle systems 
encountered in another context (see e.g. \cite{EvFoGoMu,EvKaKoMu}), including the $ABC$ model for which similar features occur \cite{ClDeEv};  
\item a tri-coupled exclusion process to represent  the stochastic 
dynamics of curves in the three-dimensional space.
\end{itemize} 
With this extended formulation, we provided a set of general conditions for
reversibility, by analyzing cycles in the state space and the corresponding invariant measure. 

This paper focuses on  non-Gibbs states and transient regimes.
In another  work in progress \cite{FaFu3}, we analyze the asymmetric simple 
exclusion process (\textsc{asep}) on a torus. Under suitable initial conditions, 
the usual sequence of empirical measures  converges in probability to a 
deterministic measure, which is the unique weak solution of a Cauchy problem. 
The method  presents some new features,  and relies on the analysis of a family 
of parabolic differential operators, involving variational calculus. This 
approach let hope for a pretty large level of generalization, and  we are 
working over its general conditions of validity. 

Sections \ref{gibbs} and \ref{nongibbs} are devoted to the stationary regime,
for which, from \cite{FaFu} and \cite{FaFu2},  the limit curves are known to
satisfy a differential system of Lotka-Volterra type which is the essence
of the fluid limits in our context. Section \ref{gibbs} solves the steady 
state regime in the reversible case. A geometric interpretation of the 
free energy is provided (involving  the algebraic area enclosed by the curve), 
as well as an urn model description for the underlying dynamical system, 
leading precisely to a Lotka-Volterra system.

Non-Gibbs states are considered in section \ref{nongibbs}. In \cite{FaFu2},
necessary and sufficient conditions for reversibility where given,
by identification of a family of independent cycles in the state graph,
for which Kolmogorov's criteria have to be fulfilled. We pursue this 
analysis by showing that irreversibility occurs as a result of particle currents
attached to these cycles. A connection between recursion properties (originating 
matrix solutions) and particle cycles in the state-graph is found, with the 
introduction of loop currents, on the analogy with electric circuits. These 
recursions at discrete level connect together invariant measures of systems of 
size $N$ (the number of sites) and of size $N-1$, and they  involve coefficients 
which are given a concrete meaning. Indeed, by means of a functional approach, we map explicitly these structure coefficients onto special constants which intervene in the Lotka-Volterra systems describing the fluid limit, as $N\to\infty$.

In the last section \ref{fluctuations}, we observe that  local equilibrium 
takes place at a rapid time-scale, compared to the diffusion time which is the natural
scale of the system. We extend the iterative scheme procedure initiated in \cite{FaFu} 
and developed in \cite{FaFu2}, which originally 
concerned only  the steady-state regime. In fact, this scheme allow us  
to express in transient regime  particle-currents in terms of  deterministic 
particle densities: this is a mere consequence of a law of large numbers.  At least when the diffusion scale is identical for all particle species,  local correlations are found 
to be absent at the hydrodynamical scale. Finally, in the spirit of the study made in \cite{BeLa}, we obtain the Lagrangian describing the fluctuations of currents, and we analyze the related Hamilton-Jacobi equations.

\section{Model definition}
\subsection{A stochastic clock model}
The system consists of an oriented path embedded in a bidimensional manifold,
with $N$ steps of equal size, each 
one being chosen among a discrete set of $n$ possible orientations, drawn from the set of angles with some given origin $\{\frac{2k\pi}{n}, k=0,\ldots,n-1\}$. The stochastic dynamics in force consists in displacing one  single point at a time without breaking the path, while keeping all  links within the set of  admissible  orientations. In this operation, two links are simultaneously displaced. This constrains quite strongly the possible dynamical rules, which are given in terms of \emph{reactions}  between consecutive links.

For any $n$, we  can define
\begin{equation}\label{exchange}
X^k X^l\ \rla_{\lambda_{lk}}^{\lambda_{kl}}\ X^l X^k,\quad
k\in[1,\,n],\, k\ne l,
\end{equation}
which in the sequel will be sometimes referred to as a local exchange process. 
It is necessary to discriminate between $n$ odd  and $n$ even. Indeed, for $n=2p$, 
there is another set of possible stochastic rules:
\begin{equation}\label{evnmod}
\begin{cases}
\DD X^k X^l\ \rla_{\lambda_{lk}}^{\lambda_{kl}}\ X^l X^k,\qquad
k=1,\ldots,n, \quad l\ne k+p,\\ 
\DD X^k X^{k+p}\ \rla_{\delta^{k+1}}^{\gamma^k} 
\ X^{k+1} X^{k+p+1},\qquad k=1,\ldots,n .
\end{cases} 
\end{equation}
The distinction is simply due to the presence, for even $n$, of \emph{folds} 
(two consecutive links with  opposite directions),  which may undergo different
transition rules, leading to a richer dynamics. The parameters $\{\lambda_{kl}\}$  represent  the exchange  rates between  two consecutive links,  while the 
$\gamma_{k}$'s  and $\delta_k$'s  correspond to the  rotation of a fold to the right or
to the left.
\subsection{Examples}\label{abcdef}
\emph{1) The simple exclusion process}

The first elementary and most studied example is the simple exclusion process, which
after mapping particles onto links corresponds to a one-dimensional fluctuating interface. In that case, we simply have a binary alphabet. Letting $X^1=\tau$ and $X^2=\bar\tau$, the reactions rewrite
\[
\tau\bar\tau \lra_{\lambda^+}^{\lambda^-}\ \bar\tau\tau,
\]
where  $\lambda^\pm$ is the transition rate for the jump of a particle to
the right or to the left. 

\emph{2) The triangular lattice and the ABC model}

Here the evolution of the random walk is restricted to the triangular lattice. 
A link (or step) of the walk is either $1$, $e^{2i\pi/3}$ or $e^{4i\pi/3}$, and 
quite naturally 
will be said to be of type A, B and C, respectively. This corresponds to the 
so-called \emph{ABC model}, since there is a coding by a 
$3$-letter alphabet. The set of \emph{transitions} (or reactions) is given by
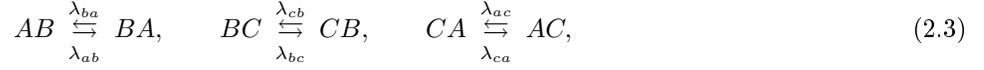
\begin{eqnarray}
AB\ \lra_{\lambda_{ab}}^{\lambda_{ba}}\ BA, \qquad 
BC\ \lra_{\lambda_{bc}}^{\lambda_{cb}}\ CB, \qquad 
CA\ \lra_{\lambda_{ca}}^{\lambda_{ac}}\ AC, \qquad 
\end{eqnarray}
where the rates are arbitrary positive numbers. Also we impose \emph{periodic
boundary conditions} on the sample paths. This model was first
introduced in \cite{EvFoGoMu} in the context of particles with
exclusion, and, for some cases corresponding to reversibility, a Gibbs form
has been found in \cite{EvKaKoMu}.

\emph {3)  A coupled exclusion model in the square lattice} 

This model was introduced in \cite{FaFu} to analyze stochastic distortions 
of a walk in the square lattice. Assuming links are counterclockwise oriented, 
the following  transitions can take place.
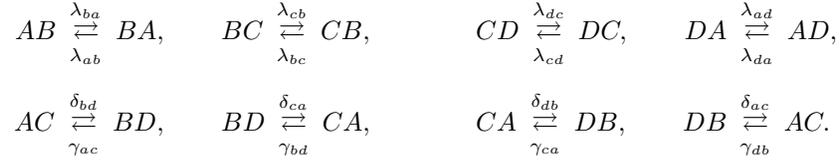
\begin{eqnarray*}
AB\ \rla_{\lambda_{ab}}^{\lambda_{ba}}\ BA, \qquad 
BC\ \rla_{\lambda_{bc}}^{\lambda_{cb}}\ CB, &\qquad& 
CD\ \rla_{\lambda_{cd}}^{\lambda_{dc}}\ DC, \qquad 
DA\ \rla_{\lambda_{da}}^{\lambda_{ad}}\ AD, \\[0.2cm] 
AC\ \rla_{\gamma_{ac}}^{\delta_{bd}}\ BD, \qquad 
BD\ \rla_{\gamma_{bd}}^{\delta_{ca}}\ CA, &\qquad& 
CA\ \rla_{\gamma_{ca}}^{\delta_{db}}\ DB, \qquad 
DB\ \rla_{\gamma_{db}}^{\delta_{ac}}\ AC. 
\end{eqnarray*}
We studied a rotation invariant version of this model, namely  when
\begin{equation}\label{symmetric}
\begin{cases}
\lambda^+ \egaldef \lambda_{ab} = \lambda_{bc} = 
\lambda_{cd} = \lambda_{da}, \\
\lambda^- \egaldef \lambda_{ba} = \lambda_{cb} = 
\lambda_{dc} = \lambda_{ad}, \\ 
\gamma^+ \egaldef \gamma_{ac} = \gamma_{bd} 
= \gamma_{ca} = \gamma_{db}. \\
\gamma^- \egaldef \delta_{ac} = \delta_{bd} 
= \delta_{ca} = \delta_{db}.
\end{cases}
\end{equation}
Define the  mapping 
$(A,B,C,D)\to(\tau^a,\tau^b)\in\{0,1\}^2$, 
such that
\begin{equation}\label{cpmapping}
\begin{cases}
A \to (0,0),\\
B \to (1,0),\\
C \to (1,1),\\
D \to (0,1).\\
\end{cases}
\end{equation}
The dynamics can be formulated in terms of coupled exclusion processes. 
The evolution of the sample path is represented by a Markov process with state space
the set of $2N$-tuples of binary random variables $\{\tau_i^a\}$ and 
$\{\tau_i^b\}$, $i=1,\ldots, N$, taking the value $1$ if a 
particle is present and $0$ otherwise. The jump rates to the right ($+$) or 
to the left ($-$) are then given by
\begin{equation}\label{taux}
\begin{cases}
\lambda_a^{\pm}(i)={\bar \tau_i^b}{\bar
\tau_{i+1}^b}\lambda^{\mp}+\tau_i^b 
\tau_{i+1}^b\lambda^{\pm} + {\bar
\tau_i^b}\tau_{i+1}^b\gamma^{\mp} + \tau_i^b{\bar
\tau_{i+1}^b}\gamma^{\pm},\\[0.2cm] \lambda_b^{\pm}(i)={\bar
\tau_i^a}{\bar \tau_{i+1}^a}\lambda^{\pm}+\tau_i^a
\tau_{i+1}^a\lambda^{\mp} + 
{\bar \tau_i^a}\tau_{i+1}^a\gamma^{\pm} +
\tau_i^a{\bar \tau_{i+1}^a}\gamma^{\mp}.
\end{cases}
\end{equation}
Notably, one sees the jump rates of a given sequence are
locally conditionally defined by the complementary sequence.

\section{Stationary regime for reversible systems}\label{gibbs}
In this section, we quote the main characteristics of the steady state distribution when 
the processes at stake are reversible.
\subsection{The general form of the invariant measure}
Up to a slight abuse in the notation, we let $X_i^k\in\{0,1\}$
denote the binary random variable representing the occupation of site
$i$ by a letter of type $k$. The state of the system is 
represented by the array $\eta\egaldef\{X_i^k, i=1,\ldots,N;
k=1,\ldots,n\}$ of size $N\times\,n$. The invariant measure of
the Markov process of interest is given by
\begin{equation}\label{inv}
\pi_\eta = 
\frac{1}{Z}\exp\bigl[-\mathcal{H}\bigl(\eta\bigr)\bigr],
\end{equation}
where
\begin{equation}\label{ansatz}
\mathcal{H}(\eta) = \frac{1}{N}
\sum_{i<j}\sum_{k,l} \alpha_{kl}\n X_i^k X_j^l,
\end{equation}
with $\alpha_{kl}\n$ and $\alpha_{lk}\n$ two $N$-dependent coefficients related by
\begin{equation}\label{alphacoef}
\alpha_{kl}\n - \alpha_{lk}\n = N\log\frac{\lambda_{kl}}{\lambda_{lk}},
\end{equation}
provided that some \emph{balance} conditions hold (see e.g. \cite{Kel}). 
For example, in  the clock model (\ref{exchange}), these conditions take the simple form
\begin{equation}\label{oddinv}
\sum_{k\ne l} \bigl(\alpha_{kl}\n -\alpha_{lk}\n\bigr) N_k = 0,
\end{equation}
and  they follow indeed directly from  Kolmogorov's criteria (applied to a particle 
crossing the system), which is tantamount to detailed balance equations.
\subsubsection{An example in the square lattice}
To show a concrete exploitation of  the form (\ref{inv}), we consider the square-lattice 
model introduced in \cite{FaFu}. It does illustrate the rules
(\ref{evnmod}). Instead of handling the problem directly with the
natural set of four letters $\{A,B,C,D\}$, we found convenient to
represent the degrees of freedom by pairs of binary components. In the
symmetric version of the model defined by (\ref{symmetric}), when 
cycles are absent ($N_a=N_b=1/2$ and $\gamma^+=\gamma^-$), we could
derive the invariant measure
\begin{equation}\label{invform}
\pi_\eta = \frac{1}{Z}\exp\bigl[\beta\sum_{i<j} 
(\tau_i^a\bar\tau_j^b - \tau_i^b\bar\tau_j^a)\bigr],
\end{equation}
with $\eta = \{(\tau_i^a,\tau_i^b),i=1\ldots N\}$  
with $\beta = \log\frac{\lambda^-}{\lambda^+}$. Let us see how this 
relates to the original formulation of the model  in terms of
the four letters $A, B, C$ and $D$.
\begin{prop}
Under the reversibility  conditions imposed on the  transitions rates
$\{\lambda_{kl},\gamma^k,\delta^k,k=1\ldots4,l=1\ldots4\}$,
the measure given by (\ref{inv})and (\ref{ansatz}) reduces to 
\begin{eqnarray}
\pi_\eta &=& \frac{1}{Z} \exp\biggl\{\frac{\beta}{2}\sum_{i<j} 
B_iA_j - A_iB_j\ +\ A_iD_j - D_iA_j \nonumber \\[0.2cm]
            &  & \qquad \quad + \ C_iB_j - B_iC_j\ +\ D_iC_j - C_iD_j\biggr\},  
            \label{invariantform}
\end{eqnarray}
and is equivalent to (\ref{invform}).
\end{prop}
The proof is not difficult, starting from  (\ref{invform}). It can also be 
achieved by  a direct argument, i.e. without using (\ref{invform}), from  
theorem 3.2 of \cite{FaFu2}.
\subsection{Free energy}
We consider again the $ABC$ model as a typical example, 
and the extension to other models will be straightforward.
Assume conditions (\ref{oddinv}) hold, so that the invariant measure 
is given by 
\[
\pi_\eta = \frac{1}{Z}\exp\bigl[\frac{1}{N}\sum_{i<j}^N\alpha_{ab}\n A_iB_j
+\alpha_{bc}\n B_iC_j+\alpha_{ca}\n C_iA_j\bigr],
\]
where the constants $\alpha_{ab}\n$, $\alpha_{bc}\n$ and $\alpha_{ca}\n$ 
take the values 
\[
\alpha_{ab}\n = N\log\frac{\lambda_{ab}}{\lambda_{ba}},\quad
\alpha_{bc}\n = N\log\frac{\lambda_{bc}}{\lambda_{cb}},\quad
\alpha_{ca}\n = N\log\frac{\lambda_{ca}}{\lambda_{ac}},
\]
while $\alpha_{ba}\n$ $\alpha_{cb}\n$ and $\alpha_{ac}\n$ are set to 
zero, to be consistent with 
(\ref{alphacoef}).  
The constraints (\ref{oddinv}) now become
\begin{equation}\label{conditions}
\frac{N_A}{N_B}  = \frac{\alpha_{bc}\n}{\alpha_{ca}\n},\quad
\frac{N_B}{N_C} = \frac{\alpha_{ca}\n}{\alpha_{ab}\n},\quad
\frac{N_C}{N_A}  = \frac{\alpha_{ab}\n}{\alpha_{bc}\n}. 
\end{equation}
Following \cite{ClDeEv}, we want to write a large deviation
functional corresponding to the above Gibbs measure when $N\to\infty$. 
Set $x=\frac{i}{N}, \,J=\exp(2i\pi/3)$, and let $Z(x)$ denote the complex number given by
\[
Z(x) = \frac{1}{N} \sum_{i=1}^{[xN]}\Bigl(\frac{A_i}{\alpha}+ 
J\frac{B_i}{\beta}+J^2\frac{C_i}{\gamma} \Bigr),
\]
where we have introduced the parameters 
\[
\alpha \egaldef \lim_{N\to\infty} \alpha_{bc}\n, \qquad 
\beta \egaldef  \lim_{N\to\infty} \alpha_{ca}\n, \qquad 
\gamma \egaldef \lim_{N\to\infty} \alpha_{ab}\n. 
\]
The sequence $\eta= \{(A_i,B_i,C_i),i=1\ldots N\}$ 
is thus represented by a discrete path  $\Gamma$ in the complex plane, 
made of oriented links having only three possible directions 
\[
\{\theta=0, \ \theta=2\pi/3,\ \theta=4\pi/3\},
\]
depending on whether a particle A, B or C is present. The length of a link 
corresponding to A, B, or C is, respectively, 
$1/(N\alpha)$,  $1/(N\beta)$ or  $1/(N\gamma)$. 

The equation of $\Gamma$ is given by a function
$Z :\egaldef x\rightarrow Z(x), x\in \mathbb{C}$. Note that condition (\ref{conditions})
ensures  $\Gamma$ is closed, that is
\[
Z(1) = \frac{1}{\alpha+\beta+\gamma}(1+J+J^2) = 0.
\]
The  area $\mathcal {A}$ enclosed by $\Gamma$ is given by
\[
\mathcal {A} \egaldef \frac{1}{2i}\oint_\Gamma\bigl(\bar z dz - z d\bar z\bigr),
\]
and, for large  $N$, this coincide with
\begin{equation} \label{AIRE}
\mathcal{A} = \frac{\sqrt{3}}{N^2} 
\sum_{l<k} \frac{A_l}{\alpha}\Bigl(\frac{B_k}{\beta} 
-\frac{C_k}{\gamma}\Bigr) + \frac{B_l}{\beta}\Bigl(\frac{C_k}{\gamma} - 
\frac{A_k}{\alpha}\Bigr) + \frac{C_l}{\gamma}\Bigl(\frac{A_k}{\alpha} 
- \frac{B_k}{\beta}\Bigr)
+ o(1). 
\end{equation}
As a result,
\[
{\cal H}(\{\eta\}) = \frac{N \alpha\beta\gamma}{2\sqrt 3}\ {\cal A} + 
\frac{3N\alpha\beta\gamma}{(\alpha+\beta+\gamma)^2} + O(1).
\]
The large deviation probability is easily obtained from the law of 
large numbers. It is given 
by
\begin{equation}\label{ldfunctional}
P_N(\rho_a,\rho_b,\rho_c) = 
\frac{1}{Z}\exp\bigl(-N{\cal F}(\rho_a,\rho_b,\rho_c)\bigr),
\end{equation}
with the free energy 
\begin{equation}\label{freen}
{\cal F}(\rho_a,\rho_b,\rho_c) = \frac{\alpha\beta\gamma}{2\sqrt{3}}\ 
{\cal A}(\rho_a,\rho_b,\rho_c)
-{\cal S}(\rho_a,\rho_b,\rho_c),
\end{equation}
where 
\[
\mathcal{A}(\rho_a,\rho_b,\rho_c) \egaldef \sqrt{3}\int_0^1dx\int_x^1dy
\frac{\rho_a(x)}{\alpha}\Bigl(\frac{\rho_b(y)}{\beta}
-\frac{\rho_c(y)}{\gamma}\Bigr)+\frac{\rho_b(x)}{\beta}\Bigl(\frac{\rho_c(y)}{\gamma}
-\frac{\rho_a(y)}{\alpha}\Bigr)+\frac{\rho_c(x)}{\gamma}\Bigl(\frac{\rho_a(y)}{\alpha}
-\frac{\rho_b(y)}{\beta}\Bigr).
\] 
and 
where the entropy term  comes from
a multinomial combinatorial factor $\frac{n!}{n_a!n_b!n_c!}$, 
namely the way of  arranging a box of  n=[N dx] sites, with $3$ species of identical 
particles having  respective populations $n_i=\rho_i(x) N dx$, $i\in\{a,b,c\}$. 
Stirling's formula for large $N$ yields
\[
{\cal S}(\rho_a,\rho_b,\rho_c) = 
- \int_0^1 dx[\rho_a(x)\log\rho_a(x)+\rho_b(x)\log\rho_b(x)+\rho_c(x)\log\rho_c(x)].
\]  

Stable and metastable  deterministic profiles correspond to local minima
of the free-energy. According to (\ref{freen}), an optimal profile is 
a compromise between a maximal entropy and  a minimum  of  the 
enclosed algebraic area. Curves of maximal entropy are typically Brownian, and they have an area which scales like $1/N$; on the other hand, the opposite extreme configuration consisting of an  equilateral triangle with negative orientation achieves the minimum algebraic area, but belongs to a class of profiles for which the entropy
contribution is equal to zero (since $\rho\log\rho$ vanishes both for $\rho=0$ and
$\rho=1$). Depending on the ratio $\alpha\beta\gamma/2\sqrt{3}$ of the two contributions, we obtain either Brownian  (the degenerate point of the deterministic equations, see below) or  deterministic profiles,  both regimes being separated by a second order phase transition. 

\subsection{Lotka Volterra systems}
Under the scaling earlier defined, letting $N\to\infty$, we show on two examples 
that the limiting invariant measure is the solution of a non-linerar differential 
system of Lotka-Volterra type.
\subsubsection{Urn model}
Consider three species, denoted by $\{A,B,C\}$, and let 
$N_a^{(N)}(t)$, $N_b^{(N)}(t)$ and $N_c^{(N)}(t)$ be the corresponding 
time-dependent populations. The system is closed, $N_a+N_b+N_c=N$.
At random times taken as exponential events, individuals do meet 
and population transfer take place at rates $\alpha$, $\beta$, $\gamma$, associated
with the reactions 
\[
\begin{cases}
\DD AB \ra_{\gamma} BB ,\\[0.2cm]
\DD BC \ra_{\alpha} CC ,\\[0.2cm] 
\DD CA \ra_{\beta}  AA  .
\end{cases}\nonumber
\]
This zero-range process is an urn-type model of Ehrenfest \emph{Class},
as defined in \cite{GoLu}, where  indivivuals, rather than urns, 
are chosen at random. When $N$ increases to infinity, we  rather consider concentrations instead of integer numbers:
\[
\rho_i(t) \egaldef \lim_{N\to\infty}\frac{N_i^{(N)}(t)}{N},
\]
for $i=a,b,c$. After a proper scaling limit, the dynamics of the 
model is described by the following  Lotka-Volterra system
\begin{equation}
\begin{cases}
\DD \frac{\partial \rho_a}{\partial x} = \rho_a(\beta \rho_c - \gamma \rho_b), \\[0.2cm]
\DD \frac{\partial \rho_b}{\partial x} = \rho_b(\gamma \rho_a - \alpha \rho_c), \\[0.2cm]
\DD \frac{\partial \rho_c}{\partial x} = \rho_c(\alpha \rho_b - \beta \rho_a),  
\end{cases}\nonumber
\end{equation}
which, after replacing $x$ by $t$ and densities by concentrations, is
nothing else but the differential system giving the invariant
measure of the $(A,B,C)$ model,  in the  fluid limit at thermodynamical
equilibrium \cite{ClDeEv}.

\subsubsection{The square lattice model}
From (\ref{invariantform}), one can write down the large deviation functional 
${\cal F}(\rho_A,\rho_B,\rho_C,\rho_D)$, [as in (\ref{ldfunctional})], together with 
the conditions ensuring an optimal profile. This leads again to a differential 
system of Lotka-Volterra class
\begin{eqnarray}\label{LotkaII}
\frac{\partial\rho_A}{\partial x} = \eta\rho_A(\rho_B - \rho_D), &\qquad&
\frac{\partial\rho_B}{\partial x} = \eta\rho_B(\rho_C - \rho_A), \nonumber \\
\frac{\partial\rho_C}{\partial x} = \eta\rho_C(\rho_D - \rho_B), &\qquad&
\frac{\partial\rho_D}{\partial x} = \eta\rho_D(\rho_A - \rho_C),  
\end{eqnarray}
in which the last equation  follows merely by summing up the three
other ones. This system is structurally different from the one
obtained in \cite{FaFu}, which involved only two independent profiles
$(\rho_a,\rho_b)$ corresponding to deterministic densities
for the particles $\tau_a$ and $\tau_b$, while in the present case there are three
($\rho_A,\rho_B,\rho_C$ for example). 

It is interesting to notice that, in both models, explicit level surfaces exist. 
Indeed, the above system satisfies $\rho_A\rho_B\rho_C\rho_D=cte$, in addition to
constraint $\rho_A+\rho_B+\rho_C+\rho_D=1$. On the other hand,
$\rho_a(1-\rho_a)\rho_b(1-\rho_b)$ is the level surface of the former
system encountered in \cite{FaFu}. This can be explained by reversing the mapping
(\ref{mapping}), so that
\begin{eqnarray}
A_i = \bar\tau_i^a\bar\tau_i^b,&\qquad& B_i = \tau_i^a\bar\tau_i^b,
\nonumber \\[0.2cm]
\label{nlmapping}C_i = \tau_i^a\tau_i^b,&\qquad&  D_i = \bar\tau_i^a\tau_i^b.   
\end{eqnarray}
This indicates that the set of $4$-tuples 
$\{\tau_i^a,\bar\tau_i^a,\tau_i^b,\bar\tau_i^b\}$ constitutes the 
elementary  blocks of the system, and that letters
$A_i,B_i,C_i,D_i$ are composite variables encoding correlations of
these building blocks. Therefore, in the continuous limit, we are left
with two different descriptions of the same system, related in a
non trivial manner. We propose now to explore more carefully this
connection. In particular, while the linear mapping 
\begin{equation}\label{mapping}
\begin{cases}
\tau_i^a = B_i+C_i , \qquad  \bar\tau_i^a = A_i+D_i, \\
\tau_i^b = C_i+D_i , \qquad  \bar\tau_i^a = A_i+B_i.  
\end{cases}
\end{equation}
still holds in the continuous limit, as a relation between
expected values
\begin{equation}\label{lmap}
\begin{cases}
\rho_a = \rho_B+\rho_C,    \\
\rho_b = \rho_C+\rho_D,
\end{cases}
\end{equation}
the non-linear equations (\ref{nlmapping}) are  instead expected to bring a 
different form, since they involve correlations.

\begin{prop}
The differential system given by 
\begin{equation}\label{stat}
\begin{cases}
\DD\frac{\partial}{\partial
x}\bigl[\log\frac{\rho^a(x)}{1-\rho^a(x)}\bigr] = 2\eta
(2\rho_b(x)-1),\\[0.3cm]
\DD\frac{\partial}{\partial
x}\bigl[\log\frac{\rho^b(x)}{1-\rho^b(x)}\bigr] = -2\eta
(2\rho_a(x)-1),
\end{cases}
\end{equation}
is related to  (\ref{LotkaII}) through the invertible functional mapping given by
\begin{equation}\label{revmap}
\begin{cases}
\rho_A = \bar\rho_a\bar\rho_b+K, \qquad  \rho_B = \rho_a\bar\rho_b-K, \\
\rho_C = \rho_a\rho_b+K, \qquad  \rho_D = \bar\rho_a\rho_b-K ,
\end{cases}
\end{equation}
where $K$ is a constant to be determined.
\end{prop}
\begin{proof}
First, let $\{\rho_B,\rho_C,\rho_D\}$
be the set of independent variables in (\ref{LotkaII}), and express them
 in terms of the new triple $\{\rho_a,\rho_b,\rho_C\}$ given by (\ref{lmap}).
 This gives
\begin{eqnarray}
\frac{\partial(\rho_a-\rho_C)}{\partial x} &=& 
\eta(\rho_a - \rho_C)(\rho_a+\rho_b -1), \nonumber \\
\frac{\partial(\rho_b-\rho_C)}{\partial x} &=& 
\eta(\rho_b - \rho_C)(1 - \rho_a+\rho_b), \nonumber \\
\frac{\partial \rho_C}{\partial x} \hspace{1cm}       
&=& \eta\rho_C(\rho_b - \rho_a). \label{rhoC}
\end{eqnarray}
Combining these equations yields
\begin{equation}\label{tmp}
\begin{cases}
\DD \frac{\partial\rho_a}{\partial x} = 
\eta\rho_a(\rho_a + \rho_b -1) +\eta\rho_C(1-2\rho_a),  \\[0.2cm]
\DD \frac{\partial\rho_b}{\partial x} = 
\eta\rho_b(1-\rho_a - \rho_b) +\eta\rho_C(2\rho_b -1),    
\end{cases}
\end{equation}
which in turn allows to express   $\rho_C$ as
\[
\rho_C = \frac{1}{\rho_a-\rho_b}\biggl(\rho_a\frac{\partial\rho_b}{\partial x}\ 
+\ \rho_b\frac{\partial\rho_a}{\partial x} \biggr).
\]

Instantiating this last value of  $\rho_C$  in (\ref{tmp}) and in (\ref{rhoC}), we obtain 
(\ref{stat}),  after immediate recombination,  together with the relation 
\[
\frac{\partial \rho_C}{\partial x} = \frac{\partial(\rho_a\rho_b)}{\partial x}.
\]
This last equation has its counterpart for $\rho_A$, $\rho_B$ and
$\rho_D$: after integration, we are left with four constants, which
reduce to the one given in (\ref{revmap}) only when compatibility
with (\ref{lmap}) is imposed.

\end{proof}

\section{Non-Gibbs steady state regime}\label{nongibbs}

We call non-Gibbs steady state regime, a regime for which the invariant measure is
not described by means of a potential. This  occurs
when reversibility is broken, that is when there exists  at least 
one cycle in the state graph
for which the Kolmogorov criteria fails. A complete set of detailed 
balanced equations cannot 
be written in such a case, there exist at least two states $\eta$ and $\eta'$,
connected  by a single particle jump, with rate $\lambda_{\eta\eta'}$, 
$\lambda_{\eta'\eta}$ such that
\begin{equation}\label{eq:gcurrent}
\lambda_{\eta\eta'} \pi_\eta - \lambda_{\eta'\eta}\pi_{\eta'}
= \phi \ne 0,
\end{equation}
if $\pi_\eta$ denotes the invariant measure.
It is the second member of this equation we wish to study in this section.
In the sequel we note $\cal S$  the state space, 
$\cal G$ the corresponding state graph, by assigning oriented edges between pair of
nodes $(\alpha,\beta)\in{\cal S}^2$, when the rate $\lambda_{\alpha\beta}$ 
is non-zero, $\cal C$ will denote a  cycle in $\cal G$ and we denote  
$\cal T$ the set of spanning trees on $\cal G$.

\subsection{The tagged particle cycle}
\begin{figure}[htb]
\begin{center}
\resizebox*{!}{4cm}{ \input{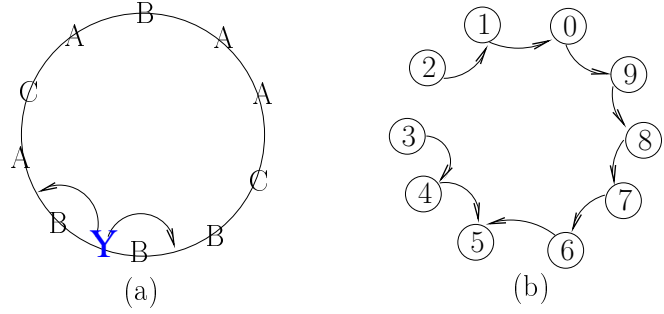}}
\caption{\label{loop}(a): relative motion of the tagged particle. (b): corresponding
state space and a spanning tree contribution to $\pi_5$}
\end{center}
\end{figure}

Cycles in the state graph for the $n$ odd model are  important in the analysis of reversibility, and they are the ones for which at least one  particle performs a complete round-trip. For example if a given particle makes $N-1$ successive jumps to the right, because of the circular geometry, the initial and final states are identical, up to a 1-step global shift to the left of the particles. As long as this particle is the only one in movement, the permutation order of the remaining other $N-1$ particles  is kept frozen.  The corresponding subsequence $\eta\nm$ will in the sequel denote  these specific cycles.
Let us examine this one particle model, by tagging a specific particle  
which is given a new label $Y$, and by following its motion conditionally
on $\eta^*=\{X_i^{k},i=1\ldots N, k\in\{1\ldots n\}\}$, the complementary frozen set of particles. This is equivalent to consider $Y$ moving in  
the inter-sites $\{i+1/2,i=0\ldots N-1\}$ of the $N-1$ frozen particles.
The question is then to analyze the steady-state regime of a 
particle moving around a circular lattice in a random environment.
To all allowed transitions which are jumps of $Y$ between sites $i-\frac{1}{2}$ and $i+\frac{1}{2}$, we let correspond the set of conditional transition rates given by
\[
\begin{cases}
\DD\lambda_y^{+}(i) = \sum_{k=1}^n \lambda_{yk}\ X_i^k,\\[0.2cm]
\DD\lambda_y^{-}(i) = \sum_{k=1}^n \lambda_{ky}\ X_i^k.
\end{cases}
\] 
Violation of condition (\ref{oddinv}) corresponds to have, 
\begin{equation}\label{def:det}
\det(\eta^*) \egaldef
\prod_{i=0}^{N-1} \lambda_y^{+}(i) - 
\prod_{i=0}^{N-1}\lambda_y^{-}(i) \ne 0,
\end{equation}
This coefficient attached to the cycle $\eta^*$ is 
the determinant of the set of flux equations
\begin{equation}\label{eq:flux}
\lambda_y^{+}(i)\pi_{i-\frac{1}{2}} - \lambda_y^{-}(i)\pi_{i+\frac{1}{2}}
= \phi(\eta^*), \quad\qquad i=0,\ldots ,N-1,
\end{equation}
giving the invariant measure $\pi_{i+\frac{1}{2}}$ which reads,
\[
\pi_{i+\frac{1}{2}} =  \frac{1}{Z} \sum_{l=1}^N
\exp\biggl\{\sum_{m=1}^{n,N}\sum_{l+1<j<i} X_j^m\log\lambda_{ym} + 
\sum_{i<j<l} X_j^m\log\lambda_{my}\biggr\}, \qquad i=0,\ldots, N-1,
\]
where $Z$ is a normalization constant. A diagramatic representation of each term in the
summation (over $l$) is given in Fig. \ref{loop}.b. Each term is in fact a spanning  tree on the reduced tagged-particle state-graph, weighted by the transitions rates and   
rooted at the considerd point ($i+\frac{1}{2}$ for $\pi_{i+\frac{1}{2}}$). 
The constant $Z$ is therefore the sum of all spanning-trees on 
the reduced tagged-particle state-graph. The probability current between site $i-\frac{1}{2}$ and site $i+\frac{1}{2}$ 
reads
\[
\lambda_y^{+}(i)\pi_{i-\frac{1}{2}} - 
\lambda_y^{-}(i)\pi_{i+\frac{1}{2}} 
= \frac{1}{Z}\biggl[\exp\bigl(\sum_{m=1}^n N_m\log\lambda_{ym}\bigr)
\ -\ \exp\bigl(\sum_{m=1}^n N_m\log\lambda_{my}\bigr)\biggr],
\]
with $N_m$ the number of particles of type $m$, a quantity independent of $i$. This 
shows
that $\phi(\eta^*)$ is a quantity  attached to the cycle $\eta^*$, 
which will be referred to as {\it cycle current} and reads
\begin{equation}\label{def:flux}
\phi(\eta^*) = \frac{1}{Z} \det(\eta^*).
\end{equation} 
Depending on the sign of $\det(\eta^*)$, the diffusion of particle 
$Y$ is biased in the right 
($\det(\eta^*)>0$) or in the left  ($\det(\eta^*)<0$) direction.
Of course the reversible case is recovered when the determinant vanishes,
which corresponds exactly to Kolmogorov's criterion.

{\bf Case of open systems: example of ASEP}

\begin{figure}[htb]
\begin{center}
\resizebox*{!}{4cm}{ \input{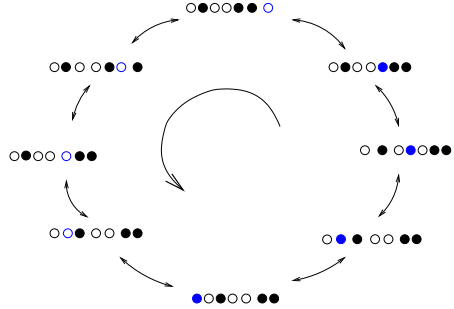}}
\caption{\label{asepcycles} Example of a tagged particle cycle in the state graph for \textsc{asep} with $7$ particles and open boundary.}
\end{center}
\end{figure}

Consider the well studied  asymmetric simple exclusion process \textsc{asep} 
with open boundary conditions, defined by $\alpha$ the rate of particle entering to the 
left side and $\beta$ the  rate at which particles exit from the right side. 
The generalization to open systems of our definition of 
the tagged particle cycle (\textsc{tpc})
is depicted in figure \ref{asepcycles}.a. We adopt the convention for the cycle 
orientation that particles move positively to the right and holes to the left.
Assume we give a tag to one of the particles. Let it perform sucessive jumps until 
reaching the right side; when it leaves the system it is in fact transformed into a hole;
We keep the tag attached to the hole which performs successive jumps in the opposite
direction  until it reaches the left 
side; again it transformed back into a particle which in turn performs jumps to the right
until the reaching of the initial position, to conclude the cycle.

\subsection{Combinatorial formulas for invariant measure  and currents}
We give here a  combinatorial way of expressing the stationary measure
on a connex finite state space $\cal S$ of size $\cal N$, the number of states. 
We consider a  continuous-time irreducible Markov
chain,  with a set of transition rates $\lambda_{\alpha\beta}$ 
between states $\alpha$ and $\beta$ and define the corresponding 
state graph $\cal G$, based on $\cal S$, by assigning oriented edges between pair of
nodes $(\alpha,\beta)$, when the corresponding rate $\lambda_{\alpha\beta}$ 
is non-zero. 
\begin{prop}\label{combin1}
The invariant measure $\pi_\alpha$ is given by 
\begin{equation}\label{invmeasure}
\pi_\alpha = \frac{\sum_{t\in{\cal T}_\alpha}w(t)}{\sum_{t\in{\cal T}}w(t)}  
\end{equation}
where $\cal T$ is the set of spanning tree over $\cal G$,
${\cal T}_\alpha$ is the set of spanning tree over $\cal G$ rooted in $\alpha$,
and $w(t)$ the weight of a tree $t$ given by
\[
w(t) = \prod_{(\alpha,\beta)\in t} \lambda_{\alpha\beta}.
\]
\end{prop}
\begin{proof}
This follows from reexpressing the solution to the steady-state equation
\[
\pi_\alpha G_{\alpha\beta} = 0,\qquad\qquad \forall\beta\in{\cal S}
\]
where G is the generator, and $G_{\alpha\beta} = -
\Bigl(\sum_\gamma \lambda_{\alpha\gamma}\Bigr)\delta_{\alpha\beta} + 
\lambda_{\alpha\beta}$, using the Cramer relation. Indeed 
since 
\[
\sum_{\beta=1}^{\cal N} G_{\alpha\beta} = 0,
\]
the set of steady-state equations is of rank ${\cal N}-1$ and $\pi_\alpha$ can 
be written as the ratio of two determinants, namely the 
cofactor $\tilde G_{\alpha {\cal N}}$ 
of $G_{\alpha {\cal N}}$ and the determinant $|\tilde G|$ 
of the matrix obtained from $G$
by replacing $G_{\beta {\cal N}}$ by $1$ for $\beta=1,\ldots, {\cal N}$.
$G$ has a structure of an admittance-matrix, as a result, expanding $\tilde G_{\alpha
\beta}$
and  $|\tilde G|$ amounts to sum over spanning trees,
\[
\tilde G_{\alpha {\cal N}} = \sum_{t\in{\cal T}_\alpha}w(t),\qquad\qquad
|\tilde G| = \sum_{t\in{\cal T}}w(t),
\]
which leads to formula (\ref{invmeasure}).
\end{proof}
From this observation we deduce a way
to express the probability currents at steady-state, which
generalizes formula (\ref{eq:flux}) and (\ref{def:flux}).
First call $\det(C)$ a coefficient attached to each cycle $C$,
\[
\det(C) \egaldef \prod_{(\gamma,\delta)\in C}\lambda_{\gamma\delta}
- \prod_{(\gamma,\delta)\in C}\lambda_{\delta\gamma},
\] 
generalizing (\ref{def:det}) and 
where the orientation of $C$ is prescribed  by the orientation of $(\alpha,\beta)$
and the product over the set $(\gamma,\delta)\in C$, is understood according
to this orientation.
Let ${\cal C}_{\alpha\beta}$ the set of  cycles
in $\cal G$ containing the oriented edge $(\alpha,\beta)$.
Let  ${\cal T}_{\cal C}$  a set of subgraph of $\cal G$, s.t. when $\cal C$ is
glued into a single node $\alpha_{\cal C}$, ${\cal T}_{\cal C}$ represents the set of 
spanning trees rooted in $\alpha_C$.
\begin{lem}
The steady state current between states $(\alpha,\beta)\in {\cal S}^2$ 
is given by
\begin{equation}\label{invcurrent}
\lambda_{\alpha\beta}\pi_\alpha - 
\lambda_{\beta\alpha}\pi_\beta = \sum_{C\in{\cal C}_{\alpha\beta}}
\frac{\sum_{t\in{\cal T}_{\cal C}}w(t)}{\sum_{t\in{\cal T}}w(t)}\det(C)
\end{equation}
\end{lem}
\begin{figure}[htb]
\begin{center}
\resizebox*{!}{5cm}{\input{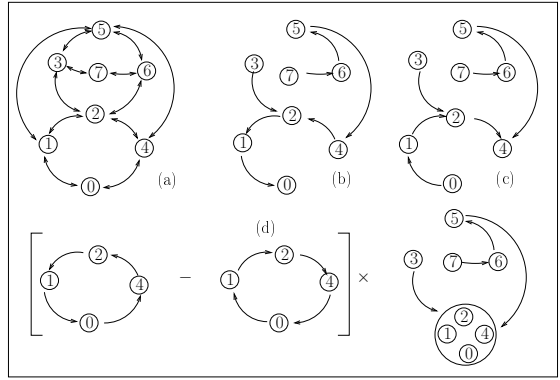}}
\caption{\label{state_graph}(a): state-graph with $N=8$ states. Arrows indicates
possible transitions. (b): a contribution to $\pi_0$. (c): a contribution
to $\pi_4$. (d): a combined contribution to $J_{14}$.}
\end{center}
\end{figure}
\begin{proof}
When $\pi_\alpha$ 
is multiplied by $\lambda_{\alpha\beta}$, each spanning tree contribution
is transformed by drawing  an oriented edge between 
$\alpha$ and $\beta$. Since the spanning tree contains by construction of 
$\pi_\alpha$ a
path going from $\beta$ to $\alpha$, the added edges
contributes to the forming of a cycle which contains $\alpha$ and $\beta$. 
If each oriented edge  in this cycle have a reversed counterpart, then in 
$\lambda_{\beta\alpha}\pi_\beta$
there is to be found a corresponding term with the same edges but with
reversed orientation in the cycle (see Fig. \ref{state_graph}).
In any case, $\det(C)$ factors out of an ensemble of contributions which consist
in drawing trees spanning all the subgraph ${\cal G}$ with end-points on $C$,
divided by the global normalization constant $\sum_{t\in{\cal T}}w(t)$. 
This complete the justification of formula (\ref{invcurrent}).  
\end{proof}
Note that $\sum_{t\in{\cal T}_{\cal C}}w(t)$ in (\ref{invcurrent}) 
represents the unormalized invariant measure of 
$\alpha_{\cal C}$ on the reduced graph ${\cal G}/{\cal C}$. This indicates 
that (\ref{invcurrent}) bears recursive properties which could be used for
asymptotic limits when the size of the system tends to infinity. 
Let us call $C$ a reversible [resp. non-reversible] cycle if $\det(C)=0$ 
[resp. $\det(C)\ne0$]. 
In the loop expansion of the currents provided by (\ref{invcurrent}), only
non-reversible cycles do contribute. For particle system, this distinction 
is embedded into a topological classification of cycles with respect to their
corresponding determinant value $\det(C)$.\\
  
{\bf Connection with the matrix ansatz for ASEP}

\begin{figure}[htb]
\begin{center}
\resizebox*{!}{8cm}{ \input{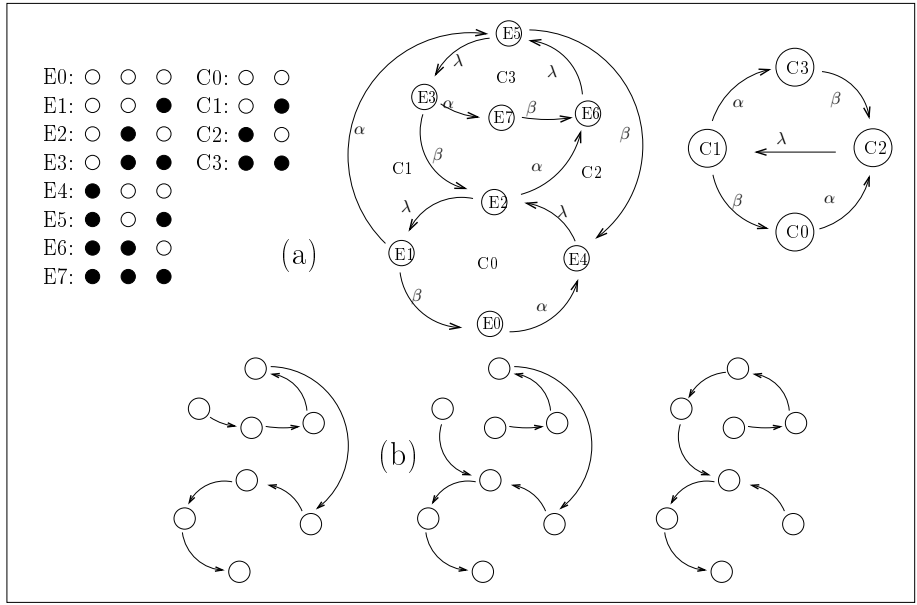}}
\caption{\label{cycleasep}(a):Graph of the state space for a \textsc{tasep}  
with three particles and the dual graph corresponding to the possibles cycles. (b)
spanning tree contributions to $\pi_{000}$.}
\end{center}
\end{figure}

For the \textsc{asep} model, a simple algorithm has been discovered \cite{DeEvHaPa}
to obtain the steady-state probabilities of each individual 
state  with the help of a matrix ansatz  . In this representation, 
a given sequence 
$\eta = 1010\ldots 00$  is  represented by a product of matrices $D$ (for $1$) 
and $E$ (for $0$), and the 
corresponding probability measure is obtained by taking the trace
\[
\pi_\eta = \frac{1}{Z}\text{Tr}\bigl(WDEDE\ldots EE\bigr),
\]
where $W$ is an additional matrix which takes into account the boundary property.
A sufficient condition for this to be the invariant measure is
that $D,E,W$ satisfy
\begin{eqnarray}
\DD \lambda_{10}DE - \lambda_{01}ED &=& D+E \label{algebra}\\[0.2cm]
\DD DW &=& \frac{1}{\beta}W \nonumber\\[0.2cm]      
\DD WE &=& \frac{1}{\alpha}W .\nonumber 
\end{eqnarray}
If $\lambda_{01}=0$, the process is totally asymmetric (\textsc{tasep}), 
particles can jump only to the right. Consider
the system with only 3 sites, which graph is depicted in figure \ref{cycleasep}.
Using these rules we find e.g. that 
\begin{align}
\pi_{000}  &= \frac{1}{Z}\alpha^3\\[0.2cm]
\pi_{100}  &= \frac{1}{Z}\Bigl(\bigl(
\frac{1}{\alpha}+\frac{1}{\beta}\bigr)\frac{1}{\lambda^2}+
\frac{1}{\alpha^2\lambda}\Bigr).
\end{align}
Comparison with the spanning tree expansion is done by counting deletions. 
A spanning tree is 
obtained from the complete graph by the deleting of a certain number of edges, 
and each deletion  is accounted for by dividing 
with respect to the corresponding transition rate.
The set of spanning trees contributing to $\pi_{000}$ is given in figure \ref{cycleasep}.b. 
The brut result (without normalization is):
\begin{align}
\pi_{000}  &\propto \Bigl(\frac{1}{\alpha\beta}+
\frac{1}{\alpha\beta}+\frac{1}{\beta\lambda} 
\Bigr)\alpha^3\\[0.2cm]
\pi_{100}  &\propto \Bigl(\frac{1}{\alpha\beta}+
\frac{1}{\alpha\beta}+\frac{1}{\beta\lambda} 
\Bigr)\Bigl(\bigl(
\frac{1}{\alpha}+\frac{1}{\beta}\bigr)\frac{1}{\lambda^2}+\frac{1}{\alpha^2\lambda}\Bigr).
\end{align}
The factor $\Bigl(\frac{1}{\alpha\beta}+\frac{1}{\alpha\beta}+\frac{1}{\beta\lambda} 
\Bigr)$ shows up for each state, and disapears after  normalization.
Nervertheless, it induces in this simple example a factor of 3 in the enumeration of terms, 
by comparison with the matrix ansatz. An underlying symmetry of the state graph is at 
the origin of this combinatorial factor.
Indeed for the \textsc{asep} system, 
the steady-state probability current between  two sequences 
$\eta$ and $\eta'$ separated by a single jump between site $i$ and $i+1$ reads,
\begin{equation}\label{asepcurrent}
\lambda_{10}\pi_{\eta}\n - \lambda_{10}\pi_{\eta'}\n = 
\pi_{\eta_i^*}\nm  +\pi_{\eta_{i+1}^*}\nm,
\end{equation}
as a consequence of (\ref{algebra}), with the subsequence $\eta_i^*$ 
[resp. $\eta_{i+1}^{*'}$] 
of $\eta$ obtained by deleting bit $i$ [resp. $i+1$].
We have not been able yet to fill the gap between (\ref{invcurrent}) and 
(\ref{asepcurrent}). We believe 
that the combinatorial arrangement which occur is due to a hierarchical 
structure of the  state-graph, revealed with the help of the tagged particle. The complete  analysis of (\ref{invcurrent}) is the subject of another work in progress. Beforehand, in the next sections, we simply propose a possible general form for the detailed current equation (\ref{eq:gcurrent}), which leads (see section \ref{hydrodyn})  to the correct form of the Lotka-Volterra equations describing the fluid limits at steady state.


\subsection{Cycle currents}
\begin{figure}[htb]
\begin{center}
\resizebox*{!}{6cm}{ \input{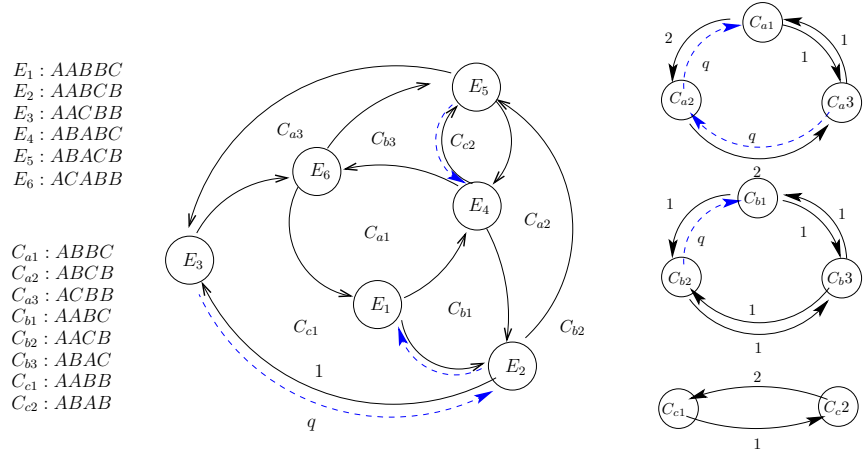}}
\caption{\label{cycleabc}Graph of the state space for a 
an asymmetric ABC model 
with five particles,
$(A,A,B,B,C)$ and the dual graph corresponding to
the possible cycles.}
\end{center}
\end{figure}

We interpret relation (\ref{asepcurrent}) in terms of {\it cycle currents}.
A transition taking place between two particles of different types, say
 $AB\rightarrow BA$, can be viewed  either as a particle $A$ travelling to the right 
or, conversely, as a particle $B$ travelling to the left.  In this exchange two joint \textsc{tpc} are involved. In the state-graph, each \textsc{tpc} defines a face,
which we will identify  with a subsequence $\eta^*$, obtained from $\eta$ by removing 
the tagged particle. Accordingly, we attach a set of variables $\{\phi(\eta^*)\}\in {\mathbb R}$ to each \textsc{tpc} face, while  currents between states are variables attached to the edges of the graph. Conservation of probability currents at a given node
is automatically fulfilled, provided that if one write (assuming a transition between 
site $i$ and $i+1$), see figure \ref{cycleabc}),
\begin{equation}\label{balance}
\lambda_{ab}\pi_{\eta}-\lambda_{ba}\pi_{\eta'} = \phi_a\bigl(\eta_i^*\bigr) - 
\phi_b\bigl(\eta_{i+1}^*\bigr),
\end{equation}
which is tantamount to changing current variables into cycle variables.

The right-hand side members in (\ref{algebra}) and   (\ref{asepcurrent}) is reminiscent of the 
second member of (\ref{balance}).
In fact we have 
\begin{align*}
\phi_a(\eta_i^*) &= \text{Tr}\bigl(W\eta_i^*\bigr) \\[0.2cm]
\phi_b(\eta_{i+1}^*) &= - \text{Tr}\bigl(W\eta_{i+1}^*\bigr)
\end{align*}

With each edge of the state-graph, we associate such an 
extended detailed balance equation. Then, eliminating all $\phi$'s from this set of equations leads to the invariant measure equation.
Consider the example given in figure \ref{cycleabc}. The transition rules are
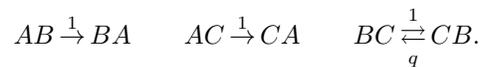
\[
AB \ra^1 BA\qquad AC\ra^1 CA \qquad BC\rla_q^1 CB.
\]
The various weights corresponding to each sequence and subsequence 
associated with cycles are given in the following 
table, for $q=0$ and $q=1$. Note that one should expect 
$\pi_{c1}=\frac{1}{3}$ and $\pi_{c2}=\frac{2}{3}$
from the subgraph of figure \ref{cycleabc}. The correction results from the 
different degeneracy w.r.t circular permutation
symmetry (4 for $C_1$ and 2 for $C_2$).    

\begin{tabular}{|c|c|c|c|c|c|c||c|c|c|c|c|c|c|c||c|c|c|}
\hline
  & $\pi_1$ & $\pi_2$ & $\pi_3$ & $\pi_4$ & $\pi_5$ 
& $\pi_6$ & $\pi_{a1}$ & $\pi_{a2}$ & $\pi_{a3}$ 
& $\pi_{b1}$ & $\pi_{b2}$ & $\pi_{b3}$ & $\pi_{c1}$ 
& $\pi_{c2}$ & $C_a$ & $C_b$ & $C_c$\\[0.2cm]
\hline
$q=0$ & $\frac{1}{10}$ & $\frac{1}{10}$ & $\frac{3}{10}$ & $\frac{1}{10}$ 
& $\frac{2}{5}$ & $\frac{2}{5}$ 
& $\frac{1}{4}$ & $\frac{1}{4}$ & $\frac{1}{2}$
& $\frac{1}{6}$ & $\frac{1}{2}$ & $\frac{1}{3}$
& $\frac{1}{2}$ & $\frac{1}{2}$ 
& $\frac{2}{5}$ & $0$ & $-\frac{1}{5}$\\[0.2cm]
\hline
$q=1$ & $\frac{1}{6}$ & $\frac{1}{6}$ & $\frac{1}{6}$ 
& $\frac{1}{6}$ & $\frac{1}{6}$ & $\frac{1}{6}$ 
& $\frac{1}{3}$ & $\frac{1}{3}$ & $\frac{1}{3}$
& $\frac{1}{3}$ & $\frac{1}{3}$ & $\frac{1}{3}$
& $\frac{1}{2}$ & $\frac{1}{2}$
& $\frac{32}{105}$ & $-\frac{41}{210}$ & $-\frac{41}{315}$\\[0.2cm]
\hline
\end{tabular}

In this two  cases one has the decomposition of the dual variables 
$\phi$ of relation \ref{balance} according to 
\begin{equation}\label{eq:structure}
\phi_x(\eta^*) = C_x\pi_{\eta^*} \qquad\text{with}\ x\in\{a,b,c\} 
\end{equation}  
with the value of the structure coefficient also given in the table. For example we have
\[
\pi_1 - q\pi_2 = C_b\pi_{b1} - C_c\pi_{c1}.
\]
This decomposition is however not valid for arbitrary $q$. 
A certain number of compatibility constraint have to be imposed on the $\phi'$,
because the \textsc{tpc} do not constitute a complete bases of cycles in the state graph.
When considering the complete system (\ref{balance}) of detailed currents, we 
have at hand  $m$  equations, $m$ being the number
of edges of the state-graph, and  $n+\nu_\text{tpc}$ unknowns,
where $n$ is the number of nodes and $\nu_\text{tpc}$ the number of \textsc{tpc}.
In matrix form, this reads
\begin{equation}\label{currents_system}
M\Pi = \Phi,
\end{equation}
where 
\begin{itemize}
\item $M$ is a $m\times n$ matrix;
\item $\Pi$ a column vector of size $n$, with the elements the invariant probability 
measure;
\item $\Phi$ is a column vector of size $m$, where each component $l$ is 
the algebraic contribution of the (two in general) \textsc{tpc}  having the 
edge corresponding to $l$ in common. 
\end{itemize}
To fix the sign conventions, we agree that  orientations of 
cycles are given by the natural orientation of the system, 
i.e. each particle travels positively from left to right. An exception is 
made for the simple exclusion system, since in this case holes travel positively 
to the left and there is only one type of \textsc{tpc}.    
\begin{align*}  
\lambda_{10}\pi_\eta - \lambda_{01}\pi_{\eta'} &= 
\phi(\eta_i^*)+\phi(\eta_{i+1}^*) \qquad\ \ \text{for}\,\textsc{ asep,}\\[0.2cm] 
\lambda_{ab}\pi_{\eta} - \lambda_{ba}\pi_{\eta'} &= 
\phi_a(\eta_i^*)-\phi_b(\eta_{i+1}^*)\qquad \text{for multi-type systems,} 
\end{align*}
(with $(i,i+1)$ the sites involved in the transition). 
From basic graph theory 
(see \cite{Berge}), the quantity giving 
the number of independent cycles in an arbitrary 
graph $\cal G$ is called the \emph{cyclomatic number}
\[
\nu({\cal G}) = m - n + p,
\]
where $n,m$ and $p$ are respectively the number of nodes, edges and components. 
In our cases, the system is  irreducible, so $p=1$. 
Since $m$ is the number of 
equations and $n+\nu_\text{tpc}$ the number
of unknown, the system is over-determined by a quantity
\[
m - (n+\nu_\text{tpc}) = \nu - \nu_\text{tpc} -1.
\]
This over-determination is understood as follows.
To each line of the matrix $M$ corresponds  a transition  between two states, 
so that  a given cycle in the state-graph corresponds to some combination of lines of 
$M$ (namely the successive transitions taking part in the cycle), and the resulting 
sub-matrix is a square matrix of size the number of states visited by the cycle. 
The corresponding determinant vanishes for all trivial cycles.
Hence the number of independent 
equations is $m-\nu+\nu_\text{tpc}$, which is equal to the number of unknown 
minus $1$, the remaining degree of freedom being related to the global 
normalization condition. 
However, a certain number of compatibility conditions have to be imposed 
on the $\phi$'s in order to eliminate safely all dependent equations of 
our system (\ref{currents_system}). These conditions are somehow related  to
the basic recurrence scheme which is at the origin of matrix-solutions
obtained in the context of \textsc{asep}, but also for multi-type
particle systems \cite{ArHeRi}. Let us see how the specific form (\ref{eq:structure})
encountered precedingly  do combine with these compatibility conditions.
\begin{lem}
The form
\begin{equation}\label{flux}
\phi_a\n(\eta^*) = C_a\n \pi\nm_{\eta^*}, 
\end{equation}
of the {\it cycle currents} fulfills the compatibility condition 
imposed by trivial cycles if and only if
\[
C_a\n C_b\nm = C_b\n C_a\nm\qquad\qquad\forall a,b\in\{1\ldots n\}.
\]
\end{lem}
\begin{proof}
Instead of proving this for an arbitrary trivial cycle, we
do it for the one depicted in figure  \ref{trivial},
the completion of the general case follows by recurrence, 
since any trivial cycle can be constructed as a combination of cycle 
of this type.
\begin{figure}[htb]
\begin{center}
\resizebox*{!}{4cm}{ \input{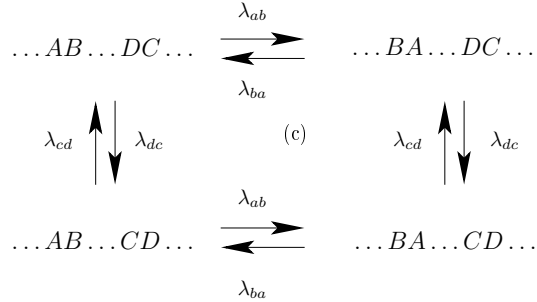}}
\caption{\label{trivial}Example of a reversible cycle.}
\end{center}
\end{figure}
To fix  some notation, let $\eta^1$,
$\eta^2$, $\eta^3$ and $\eta^4$ be the states visited by the cycle, 
with $i$ the position of  $A$ and $j$ the position of $C$  in $\eta_1$, so that  
\begin{align*}
\eta^1 &= \ldots{\bf AB}\ldots{\bf CD}\ldots
\qquad \eta_i^{1*}=\ldots{\bf B}\ldots{\bf CD}\ldots
\qquad \eta_{i+1}^{1*}=\ldots{\bf A}\ldots{\bf CD}\ldots\\[0.2cm]
\eta^2 &= \ldots{\bf BA}\ldots{\bf CD}\ldots
\qquad \eta_j^{2*}=\ldots{\bf BA}\ldots{\bf D}\ldots
\qquad \eta_{j+1}^{2*}=\ldots{\bf BA}\ldots{\bf C}\ldots\\[0.2cm]
\eta^3 &= \ldots{\bf BA}\ldots{\bf DC}\ldots
\qquad \eta_i^{3*}=\ldots{\bf A}\ldots{\bf DC}\ldots
\qquad \eta_{i+1}^{3*}=\ldots{\bf B}\ldots{\bf DC}\ldots\\[0.2cm]
\eta^4 &= \ldots{\bf AB}\ldots{\bf DC}\ldots
\qquad \eta_j^{4*}=\ldots{\bf AB}\ldots{\bf C}\ldots
\qquad \eta_{j+1}^{4*}=\ldots{\bf AB}\ldots{\bf D}\ldots
\end{align*}
The sub-system of \ref{currents_system} corresponding to this cycle reads
\begin{align*}
\lambda_{ab}\pi_{\eta^1} - \lambda_{ba}\pi_{\eta^2} 
&= \phi_a[\eta_i^{1*}] - \phi_b[\eta_{i+1}^{1*}],\qquad(a)\\[0.2cm]
\lambda_{cd}\pi_{\eta^2} - \lambda_{dc}\pi_{\eta^3} 
&= \phi_c[\eta_j^{2*}] - \phi_d[\eta_{j+1}^{2*}],\qquad(b)\\[0.2cm]
\lambda_{ba}\pi_{\eta^3} - \lambda_{ab}\pi_{\eta^4} 
&= \phi_b[\eta_i^{3*}] - \phi_a[\eta_{i+1}^{3*}],\qquad(c)\\[0.2cm]
\lambda_{dc}\pi_{\eta^4} - \lambda_{cd}\pi_{\eta^1} 
&= \phi_d[\eta_j^{4*}] - \phi_c[\eta_{j+1}^{4*}].\qquad(d) 
\end{align*}
As already noted, these equations are not independent. Hence the combination
$\lambda_{cd}(a)+\lambda_{ba}(b)+\lambda_{dc}(c)+\lambda_{ab}(d)$ eliminates one 
equation, but with the resulting constraint on the $\phi$'s:
\begin{align}\label{subid}
&\lambda_{cd}\phi_a[\eta_i^{1*}]-\lambda_{dc}\phi_a[\eta_{i+1}^{3*}]\ +\ 
\lambda_{dc}\phi_b[\eta_i^{3*}]-\lambda_{cd}\phi_b[\eta_{i+1}^{1*}]\ +\nonumber\\[0.2cm]
&\lambda_{ba}\phi_c[\eta_j^{2*}]-\lambda_{ab}\phi_c[\eta_{j+1}^{4*}]\ +\ 
\lambda_{ab}\phi_d[\eta_j^{4*}]-\lambda_{ba}\phi_d[\eta_{j+1}^{2*}]\ =\ 0.
\end{align}
$\eta_i^{1*}$ and $\eta_{i+1}^{3*}$
are in correspondence through the transition $CD\to DC$ at site $j$, $j+1$,
as well as $\eta_j^{2*}$ and $\eta_{j+1}^{4*}$ with respect to the transition
$AB\to BA$ at site $i,i+1$ \ldots. From the hypothesis of the lemma, (\ref{subid}) rewrites
\begin{align*}
&C_a\n \left(C_c\nm \pi\nmm_{\eta_{i,j}^{1**}}\ -\ 
C_d\nm \pi\nmm_{\eta_{i,j+1}^{1**}}\right)\ +\ 
C_b\n \left(C_d\nm \pi\nmm_{\eta_{i,j}^{3**}}\ -\ 
C_c\nm \pi\nmm_{\eta_{i,j+1}^{3**}}\right)\ +\ \\[0.2cm]
&C_c\n \left(C_b\nm \pi\nmm_{\eta_{i,j}^{2**}}\ -\  
C_a\nm \pi\nmm_{\eta_{i+1,j}^{2**}}\right)\ + 
C_d\n \left(C_a\nm \pi\nmm_{\eta_{i,j}^{4**}}\ -\  
C_b\nm \pi\nmm_{\eta_{i+1,j}^{4**}}\right)\ =\ 0,
\end{align*}
where $\eta_{i,j}^{1**}$ is the sequence obtained from $\eta^1$ by suppressing letters
at site $i$ and $j$.The elimination of letters in sequences is a commutative process, 
therefore this last equality holds because of the following identities: 
\[
\eta_{i,j}^{1**}=\eta_{i+1,j}^{2**},\quad \eta_{i,j}^{3**}=\eta_{i+1,j}^{4**},\quad
\eta_{i,j}^{2**}=\eta_{i,j+1}^{3**}, \quad \eta_{i,j}^{4**}=\eta_{i,j+1}^{1**}.
\]
\end{proof}
The complete study to establishing the range of validity of the recurrence relation
(\ref{balance}) altogether with (\ref{flux})  is the object of another work in progress. We expect that in general this relation to be valid only asymptotically for large $N$, which could be proved possibly by selecting the dominant terms in the expansion (\ref{invcurrent}).

\subsection{Fluid limits}\label{hydrodyn}
In this section we examine how the  microscopic coefficients $C_k\n$,  whenever  
(\ref{flux}) holds, can be transposed at macroscopic level and how they are related to important coefficients showing up in the Lotka-Volterre equations of the fluid limit. Using the preliminary study \cite{FaFu3}, where a new functional method was introduced to handle the hydrodynamic limit of a simple exclusion process, we consider hereafter the $n$-type case. 

\subsubsection{Functional approach}
Let $\phi_k, k=1\ldots n$ a set of arbitrary functions in ${\bf C}^2[0,1]$, 
$\Gb\n \egaldef \ZZ/N\ZZ$ the discrete torus (circle).
For $i\in\Gb\n$, $X_i^k(t)$ is a binary random variable and, at time $t$, 
the presence of a particle of type $k$ at site $i$ is equivalent to $X_i^k(t)=1$. 
The exclusion constraint reads 
\[
\sum_{k=1}^n X_i^k(t) =1 , \quad\forall i\in{\bf G}.
\] 
The whole trajectory is represented by 
$\eta\n(t)\egaldef\{X_i^k(t),i\in{\bf G}\n,k=1\ldots n\}$ which  is a Markov process.  
$\Omega\n$ will denote its generator and ${\cal F}_t\n = \sigma(\eta\n(s),s\le t)$ 
is the associated natural filtration. 

Define the real-valued positive measure 
\[
Z_t\n[\phi] \egaldef \exp\left[\frac{1}{N}\sum_{k=1,i\in{\bf G}\n}^n 
\phi_k\bigl(\frac{i}{N}\bigr)X_i^k\right],
\]
where $\phi$ denotes the set $\{\phi_k,k=1\ldots n\}$.
In \cite{FaFu3} the convergence of this measure was analyzed for $n=2$. 
A functional integral operator was used to characterize limit points of this measure, 
these were shown to be indeed the unique weak solution of a partial differential 
equation of Cauchy type.    

In what follows, we will be interested in the  quantities
\[
\begin{cases}
f_t\n(\phi) \egaldef 
\Bigl[{\mathbb E}\bigl(Z_t\n[\phi]\bigr)\Bigr], \\[0.3cm]
g_t\n(\phi) \egaldef 
\log\Bigl[{\mathbb E}\bigl(Z_t\n[\phi]\bigr)\Bigr],
\end{cases}
\]
respectively  the moment and cumulant generating function.
The idea of using  $Z_t\n[\phi]$  is that the generator, 
when applied to $Z_t\n$,  can be expressed as a differential operator with respect to 
the arbitrary functions $\phi$. Indeed, we have
\[
\Omega\n\bigl[Z_t\n\bigr] = L_t\n Z_t\n,
\]
with
\[
L_t\n = N^2\sum_{k\ne l,i\in{\cal G}\n}\tilde\lambda_{kl}
\frac{\partial^2}{\partial\phi_k(\frac{i}{N})\partial\phi_l(\frac{i+1}{N})},
\]
after having set
\[
\begin{cases}
\DD \Delta\psi_{kl}\bigl(\frac{i}{N}\bigr) \egaldef 
\phi_k\bigl(\frac{i+1}{N}\bigr) - \phi_k\bigl(\frac{i}{N}\bigr)
+\phi_l\bigl(\frac{i}{N}\bigr) - \phi_l\bigl(\frac{i+1}{N}\bigr) , \\[0.4cm]
\DD \tilde\lambda_{kl}(i,N) \egaldef 2\lambda_{kl}(N)
e^{\frac{\Delta\psi_{kl}(\frac{i}{N})}{2N}}
\sinh\Bigl(\frac{\Delta\psi_{kl}\bigl(\frac{i}{N}\bigr)}{2N}\Bigr).
\end{cases}
\]
We introduce now  the key quantities for hydrodynamic scalings, by assuming
an  asymptotic expansion of the form
\[
\lambda_{kl}(N) = D\bigl(N^2 + \frac{\alpha_{kl}}{2}N\bigr) + 
{\cal O}(1),\qquad\forall k,l\ k\ne l,
\]
where $\alpha_{kl}=-\alpha_{lk}$ are real constants. Here the system is assumed 
to be \emph{equidiffusive}, 
which means there exists a constant $D$ such that, for all pairs $(k,l)$,
\[
\lim_{N\to\infty}\frac{\lambda_{kl}(N)}{N^2} = D.
\]
From now on we will omit the argument of $\lambda_{kl}(N)$ and retain the 
initial notation $\lambda_{kl}$. 
The coefficients $\alpha_{kl}$ express the asymmetry between types $k$ and $l$.
Now one can write 
\begin{align}\label{geneq}
\frac{\partial f_t\n}{\partial t} = 
N^2\sum_{k\ne l,i\in{\bf G}\n}^n
\tilde\lambda_{kl}(i,N)
\frac{\partial^2 f_t\n}{\partial\phi_k(\frac{i}{N})
\partial\phi_l(\frac{i+1}{N})} .
\end{align}
To rearrange the sum in (\ref{geneq}), in order to select dominant terms in 
the expansion with respect to $1/N$, we make use of the exclusion property,  
which is formally equivalent to
\[
\sum_{k=1}^n  \frac{\partial}{\partial\phi_k(\frac{i}{N})} = \frac{1}{N}.
\] 
Since we are on the circle $i\in\Gb\n$, Abel's summation  formula does not produce any 
boundary term, so that, skipping  details, (\ref{geneq}) can be rewritten as
\begin{align}
\frac{\partial f_t\n}{\partial t} =
D N^2 &\sum_{k=1,i\in{\bf G}\n}^n \Bigl[\phi_k\bigl(\frac{i+1}{N}\bigr)
-\phi_k\bigl(\frac{i}{N}\bigr)\Bigr]
\Bigl[\frac{\partial f_t\n}{\partial\phi_k(\frac{i}{N})}
-\frac{\partial f_t\n}{\partial\phi_k(\frac{i+1}{N})}
\nonumber\\[0.2cm]
&+\frac{1}{2}\sum_{l\ne k}\alpha_{kl}
\Bigl(\frac{\partial^2 f_t\n}{\partial\phi_k(\frac{i}{N})
\partial\phi_l(\frac{i+1}{N})} 
+\frac{\partial^2 f_t\n}{\partial\phi_l(\frac{i+1}{N})
\partial\phi_k(\frac{i}{N})}\Bigr)\Bigr]+{\cal O}(N^{-1}).
\end{align}
It is worth remarking that operators like $\DD\frac{\partial}{\partial\phi_k(\frac{i}{N})}$ 
and $\phi_k(\frac{i+1}{N})-\phi_k(\frac{i}{N})$ produce a scale factor $1/N$, 
while $\DD\frac{\partial}{\partial\phi_k(\frac{i}{N})} - 
\frac{\partial}{\partial\phi_k(\frac{i+1}{N})}$ and 
$\DD\frac{\partial}{\partial\phi_k(\frac{i+1}{N})}
\frac{\partial}{\partial\phi_l(\frac{i}{N})}$ scale as $1/N^2$: this explains 
the selection of dominant terms in the above expansion.
 
Let $N\to\infty$ and assume the convergence of the sequence $f_0\n$. Then, 
from the tightness  of the process, together with a zeste of variational and 
complex variable calculus, as in \cite{FaFu3}, we claim [the proof is omitted]  
$f_t\n$ also converges, in a 
\emph{good tempered}  functional space, and its limit $f_t$ satisfies
the functional  integral equation 
\[
\frac{\partial f_t}{\partial t} = D\int_0^1 dx \sum_{k=1}^n\phi_k(x)
\ \frac{\partial}{\partial x}\Bigl[\ \frac{\partial}{\partial x}
\frac{\partial f_t}{\partial\phi_k(x)}
-\sum_{l\ne k}\alpha_{kl}\Bigl(\frac{\partial^2 f_t}{\partial\phi_k(x)
\partial\phi_l(x)}\Bigr)\Bigr].
\]
Similarly, the cumulant characteristic function is a solution of
\begin{align}\label{eqcar}
\frac{\partial g_t}{\partial t} = D\int_0^1 dx \sum_{k=1}^n\phi_k(x)
&\ \frac{\partial}{\partial x}\Bigl[\ \frac{\partial}{\partial x}
\frac{\partial g_t}{\partial\phi_k(x)}
\nonumber\\[0.2cm]
&-\sum_{l\ne k}\alpha_{kl}\Bigl(\frac{\partial g_t}{\partial\phi_k(x)}
\frac{\partial g_t}{\partial\phi_l(x)} 
-\frac{\partial^2 g_t}{\partial\phi_k(x)
\partial\phi_l(x)}\Bigr)\Bigr].
\end{align}
Assume  at time $0$ the given  initial  profile  $\rho_k(x,0)$ to be twice 
differentiable with repect to $x$. Then (\ref{eqcar}) is given by
\[
g_t\bigl(\phi\bigr) = \int_0^1 dx \sum_{k=1}^n\rho_k(x,t)\phi_k(x),
\]
where $\rho_k(x,t)$ satisfy  the hydrodynamic system 
of coupled Burger's equations 
\[
\frac{\partial\rho_k}{\partial t} = D\Bigl[\frac{\partial^2\rho_k}{\partial x^2}
+ \frac{\partial}{\partial x}\Bigl(\sum_{l\ne k}\alpha_{lk}\rho_k\rho_l\Bigr)\Bigr]
,\qquad k=1,\ldots, n,
\]
with a set of given initial conditions $\rho_k(x,0), k=1,\ldots,n$. 

\textbf{Remark} It is important to note that, without  differentiability conditions 
for the intial profiles $\rho_k(x,0)$, one can only assert the existence of 
\emph{weak solutions} (in the sense of Schwartz's distributions) of Burger's system.

\subsubsection{Functional equation at steady-state}
\begin{theo}
Consider  a  particle system  of size $N$, with rules \ref{exchange}, with  
$n$ types of particles and periodic boundary conditions. Assume the detailed 
current equations holds, for any pair of particle types $k$ and $l$,
\begin{equation}\label{structure}
\lambda_{kl}\n \pi_\eta - \lambda_{lk}\n \pi_{\eta'} =  C_k\n \pi\nm_{\eta_i^*}
- C_l\n \pi\nm_{\eta_{i+1}^*},\qquad k,l=1\ldots n,
\end{equation}
Then the limit functional 
$\DD f_\infty[\phi] = \lim_{N\to\infty} f_\infty\n[\phi]$,
where
\[
f_\infty\n[\phi] = \sum_{\{\eta\}} \pi_\eta\exp\Bigl(\frac{1}{N}\sum_{k=1,i=1}^{n,N}
X_i^k\phi_k(\frac{i}{N})\Bigr),
\]
satisfies the equation
\begin{equation}\label{fluide}
\frac{\partial}{\partial x}\frac{\partial f_\infty}{\partial\phi_k(x)}
\ +\ \sum_{l\ne k}\alpha_{kl}
\frac{\partial^2 f_\infty}{\partial\phi_k(x)\partial\phi_l(x)}
\ =\ 
c_k f_\infty\ - v\frac{\partial f_\infty}{\partial\phi_k(x)},
\end{equation}
under the \emph{fundamental scaling} 
\[
\lim_{N\to\infty} \log\frac{\lambda_{kl}\n}{\lambda_{lk}\n} = 
\alpha_{kl}\quad \text{and}\quad\forall l\ne k,\ 
\lim_{N\to\infty} \frac{N^2 C_k\n}{\lambda_{kl}\n}= \lim_{N\to\infty}\frac{C_k\n}{D}  = c_k, 
\]
with 
\[
v \egaldef \sum_{l=1}^n c_k.
\]
\end{theo}

\begin{proof}
We use the notation of  section \ref{hydrodyn}. In order to extract additional 
information at steady state, we refine our preceding variational analysis by 
defining the functional
\begin{equation}\label{composite}
T\n\bigl(\{\phi,\partial_x\phi\}\bigr)
= \frac{N^2}{2}\Bigg[\sum_{k\ne l,i=1}^{n,N}
\tilde\lambda_{kl}\n
\frac{\partial^2}{\partial\phi_l(\frac{i}{N})
\partial\phi_k(\frac{i+1}{N})}\\[0.2cm]
+\tilde\lambda_{lk}\n
\frac{\partial^2}{\partial\phi_k(\frac{i+1}{N})
\partial\phi_l(\frac{i}{N})}\Bigg]
f_\infty\n,
\end{equation}
which corresponds to the second member of equation (\ref{geneq}) at steady state,
and where it is understood that the sets $\{\phi\}\egaldef 
\{\phi(\frac{i}{N}),i=1\ldots N\}$ and
$\{\partial\phi\}\egaldef \{\frac{\partial\phi}{\partial x}(\frac{i}{N}),i=1\ldots N\}$
are taken as independant parameters.
This functional can be writen in two different manners.
Recalling the definitions
\[
\bigtriangleup\psi_{kl}(i) \egaldef \phi_k(i+1)-\phi_k(i)-\phi_l(i+1)+\phi_l(i)
\egaldef\bigtriangleup\psi_k(i) - \bigtriangleup\psi_l(i),
\]
(\ref{composite}) may be rewritten in the form
\begin{align}
T\n\bigl(\{\phi,\partial_x\phi\}\bigr) =
N D &\sum_{k=1,i=1}^{n,N}\partial_x\phi_k(\frac{i}{N})
\Bigl[\frac{\partial f_\infty^{(N)}}{\partial\phi_k(\frac{i}{N})}
-\frac{\partial f_\infty^{(N)}}{\partial\phi_k(\frac{i+1}{N})} \nonumber
\\[0.2cm]
&+\sum_{l\ne k}\frac{\alpha_{kl}}{2}
\Bigl(\frac{\partial^2 f_\infty^{(N)}}{\partial\phi_k(\frac{i}{N})
\partial\phi_l(\frac{i+1}{N})}
+\frac{\partial^2 f_\infty^{(N)}}{\partial\phi_k(\frac{i+1}{N})
\partial\phi_l(\frac{i}{N})}\Bigr)\Bigr] +{\cal O}(\frac{1}{N}), \label{hydro1}
\end{align}
On the other hand, combining the sums in (\ref{composite}) yields \begin{align}
T\n\bigl(\{\phi,\partial_x\phi\}\bigr) = N^2\sum_{k,l=1,i=1}^{n,N}\sum_{\{\eta\}}
&e^{\frac{1}{N}\vec\phi.\vec\eta+\frac{1}{2N}\bigtriangleup\psi_{kl}(i)}
\sinh\frac{\bigtriangleup\psi_{kl}(i)}{2N} \nonumber\\[0.2cm]
&\times X_i^kX_{i+1}^l\Bigl[\lambda_{kl}\n \pi\n_\eta -\lambda_{lk}\n \pi\n_{T_i\eta} \Bigr],
 \label{hydro2}
\end{align}
where $\eta$ is a given configuration, $T_i\eta$ being the one obtained from
$\eta$ by exchanging  $i$ and $i+1$, and the shorthand notation
\[
\vec\phi.\vec\eta = \sum_{k=1,i=1}^{n,N} X_i^k\phi_k\bigl(\frac{i}{N}\bigr).
\]
 From  the assumptions in the statement of the proposition,  we can 
rewrite (\ref{hydro2}) as
\begin{align}
T\n\bigl(\{\phi,\partial_x\phi\}\bigr) = N^2\sum_{k,l=1,i=1}^{n,N}\sum_{\{\eta\}}
&e^{\frac{1}{N}\vec\phi.\vec\eta+\frac{1}{2N}\bigtriangleup\psi_{kl}(i)}
\sinh\frac{\bigtriangleup\psi_{kl}(i)}{2N} \nonumber\\[0.2cm]
&\times X_i^kX_{i+1}^l\Bigl[
C_k\n \pi\nm_{\eta_i^*}
-C_l\n \pi\nm_{\eta_{i+1}^*}
\Bigr], \label{hydro3}
\end{align}
where $\eta_i^*$ is the sequence obtained from $\eta$ by removing the site $i$.
 We also have
\[
\sum_{\{\eta\}} X_i^k \pi_{\eta_i^*}
e^{\frac{1}{N}\vec\phi.\vec\eta} = f_\infty\nm[\phi_i^*]
e^{\frac{1}{N}\phi_k(\frac{i}{N})},
\]
where $f_\infty\nm[\phi_i^*]$ means that $f_\infty\nm$ is considered as a  
function of the $n(N-1)$ variables 
$\{\phi_k(\frac{j}{N}), k=1,\ldots, n; j=1,\ldots, N,j\ne i\}$.
Using all these ingredients,  expanding (\ref{hydro3}) in powers of 
$\frac{1}{N}$ and keeping the dominant terms,  we get
\begin{equation}\label{hydro4}
T\n\bigl(\{\phi,\partial_x\phi\}\bigr)  = \frac{N^2}{2}\sum_{k,l=1,i=1}^{n,N}
\bigtriangleup\psi_{kl}(i)
\Bigl[
C_k\n \frac{\partial f_\infty\nm}{\partial\phi_l(\frac{i+1}{N})}
-C_l\n \frac{\partial f_\infty\nm}{\partial\phi_k(\frac{i}{N})}
\Bigr]+\Oc\Bigl(\frac{1}{N}\Bigr) .
\end{equation}
Now, rearranging the summation, using the exclusion property
\[
\sum_{l=1}^n\frac{\partial}{\partial\phi_l(\frac{i}{N})}=\frac{1}{N},
\]
comparing (\ref{hydro1}) and (\ref{hydro4}), we  finally obtain
\begin{align*}
N^2\sum_{k=1,i=1}^{n,N}\partial_x\phi_k\bigl(\frac{i}{N}\bigr)
&\biggl[\frac{\partial f_\infty^{(N)}}{\partial\phi_k(\frac{i}{N})}
-\frac{\partial f_\infty^{(N)}}{\partial\phi_k(\frac{i+1}{N})}
+\frac{\alpha^{kl}}{2}
\Bigl(\frac{\partial^2 f_\infty^{(N)}}{\partial\phi_k(\frac{i}{N})
\partial\phi_l(\frac{i+1}{N})}
+\frac{\partial^2 f_\infty^{(N)}}{\partial\phi_k(\frac{i+1}{N})
\partial\phi_l(\frac{i}{N})}\Bigr)\biggr] \\[0.2cm]
&= N^2\sum_{k,i=1}^{n,N}\partial_x\phi_k\bigl(\frac{i}{N}\bigr)
\Bigl[\frac{C_k\n}{D} f_\infty\nm -\sum_{l=1}^n \frac{C_l\n}{D}
\frac{\partial f_\infty\nm}{\partial\phi_k(\frac{i}{N})}\Bigr] + \Oc\Bigl(\frac{1}{N}\Bigr).
\end{align*}
As the last equality holds for any $\partial_x\phi_k$, letting $N\to\infty$  implies  
easily (\ref{fluide}), which was to be proved.
\end{proof}

\subsubsection{Lotka-Volterra systems and out-of-equilibrium stationary states}
Here we will make the link between the structure coefficients 
of the current equations (\ref{structure})  and the fluid limit description of stationary states. A solution is sought of the form 
\[
f_\infty(\phi) = \exp\Bigl(\int_0^1dx\sum_{k=1}^N\rho_k^\infty(x)\phi_k(x)\Bigr),
\] 
which, instantiated into (\ref{fluide}), yields gives the following equations for the 
$\rho_k^\infty$'s .
\[
\frac{\partial\rho_k^\infty}{\partial x}-
\rho_k^\infty\sum_{l\ne k}\alpha^{kl}\rho_l^\infty
= c_k - v\rho_k^\infty,\qquad k=1\ldots n .
\]
The interpretation of this system is now quite clear : it is exactly  a particular 
stationary solution  of the system formed by the coupled Burger's equations
\[
\frac{\partial\rho_k}{\partial t} = \frac{\partial^2\rho_k}{\partial x^2}
-\frac{\partial}{\partial x}\Bigl(\rho_k\sum_{l\ne k}\alpha^{kl}\rho_l\Bigr),
\qquad k=1\ldots n,
\]
where the functions $\rho_k$ are sought in the class 
\[
\rho_k(x,t) \egaldef  \rho_k^\infty(x-vt),
\]
the variable $(x-vt)$ being taken [modulo~$1$]. Hence, there is a frame rotating 
at velocity $v$, in which $\rho_k^\infty$ is periodic. Moreover, in this frame, 
the stationary currents do not vanish and have constant values
\[
J_k(x) =  
\frac{\partial\rho_k^\infty}{\partial x}+
\rho_k^\infty\bigl(v-\sum_{l\ne k}\alpha^{kl}\rho_l^\infty\bigr) = c_k.
\]
Therefore, while the macroscopic constants $\{c_k,k=1,\ldots ,n\}$ 
are in principle determined from the periodic boundary conditions constraints and 
from the fixed average values of each particle species, they can also be directly 
derived  from the microscopic model.


\section{Transient regime and fluctuations}\label{fluctuations}
The goal of this section is twofold : first, establish relationships between currents
and particle densities at the deterministic level by means of the law of large 
numbers; secondly, compute the stochastic corrections to these relationships 
for large but finite systems by using central limit theorems and large deviations. 

\subsection{Time-scale for local equilibrium}
In keeping with our approach, we discuss the question of local equilibrium \cite{Sp}
by means  of the following functional
\[
Y_t\n \egaldef \exp\biggl[\frac{1}{N}\sum_{k,l=1,i=1}^{n,N} 
\phi_{kl}(\frac{i}{N})X_i^kX_{i+1}^l \biggr].
\]
Without entering into cumbersome technical details, let us just notice that 
the explicit computation of  $L_t\n Y_t\n$ shows that 
$L_t\n Y_t\n$ scales like ${\cal O}(N)$
instead of ${\cal O}(1)$ as  $L_t\n Z_t\n$. This fact can be interpreted as follows.
The empirical measure 
\[
\mu_t\n \egaldef \frac{1}{N}\sum_{k,l=1,i=1}^{n,N} 
\phi_{kl}(\frac{i}{N})X_i^kX_{i+1}^l
\] 
is a convolution of the distribution of interfaces between particle domains with 
a set of arbitrary functions. To any given particle density distribution, drawn from 
the set of local hydrodynamic  densities, there corresponds an arrangement of these 
interfaces which somehow characterizes the local correlations between particles. 
At steady-state, at least in the reversible case, it is easy to  show that these
correlations vanish. Moreover this scaling tells us that correlations 
vanish at a time-scale  \emph{faster} than the diffusion scale, by a factor of $N$. 
Therefore, even in transient regime, we expect  correlations to be  negligible 
for the family of diffusive processes under study. A more formal proof of this fact is postponed to the completion of the functional approach initiated in \cite{FaFu3}. 

\subsection{Hydrodynamical currents}\label{hydrocurrents}
In our preceding studies, we devised  a scheme to obtain a fluid limit at steady state, 
first for the reversible square-lattice model in \cite{FaFu}, and also for the non-reversible
\textsc{abc} model  \cite{FaFu2}. Here we generalize this procedure to transient $n$-type particle systems, resting upon the hydrodynamic hypothesis, which will be precisely stated. The principle of the method is to reverse the relationship between particle and current variables in a suitable manner, in order to  apply a law of large numbers.

\subsubsection{Diffusion models}\label{scheme2}
The system corresponds to rules (\ref{exchange}).
For any particle-type $k$, the rescaled discrete current reads 
\begin{equation}\label{dcurrent}
J_k\n\bigl(\frac{i}{N}\bigr) \egaldef
\lambda_k^+(i+1)X_i^k
- \lambda_k^-(i)X_{i+1}^k,\qquad i=1,\ldots, N,
\end{equation}
with
\[
\begin{cases}
\DD \lambda_k^+(i) 
\egaldef \sum_{l\ne k} \frac{\lambda_{kl}}{N}X_i^l +\Gamma_kX_i^k,\\[0.2cm]
\DD \lambda_k^-(i) 
\egaldef \sum_{l\ne k} \frac{\lambda_{lk}}{N}X_i^l +\Gamma_kX_i^k,
\end{cases}
\]
where arbitrary  constants $\Gamma_k$ have been introduced (they not 
modify the value of $J_k$) to ensure that the  $\lambda_k^\pm$'s never 
vanish. To be consistent with other scalings,   $\Gamma_k$ is assumed to scale like $N
$.  
Our hypothesis is that $J_k$ has a limiting distribution, 
$J_k(x)$, such that, for any integrable complex-valued  function $\alpha$,
\begin{equation}\label{hhydro}
\lim_{N\to\infty} \frac{1}{N}\sum_{i=1}^N\alpha\bigl(\frac{i}{N}\bigr)
J_k\n\bigl(\frac{i}{N}\bigr)
=\int_0^1\alpha(x)J_k(x)dx.
\end{equation}
In addition, the system will be said {\it equidiffusive}, if there exists 
a single diffusion constant $D$,  such that, for all pair of species $(k,l)$, 
\[
\lim_{N\to\infty}\frac{\lambda_{kl}(N)}{N^2} = D  \qquad \mathrm{[equidiffusion]} .
\]
To simplify the notation, consider equation
for $k=1$, writing $J_a\egaldef J_1$ and replacing $X_i^1$ by $A_i$. Then solving 
 (\ref{dcurrent}) as a linear system yields
\[
A_{i+1} = \frac{\lambda_a^+(i+1)A_i - J_a\n\bigl(\frac{i}{N}\bigr)}
{\lambda_a^-(i)}.
\]
This  relationship between $A_i$ and $A_{i+1}$ can be 
iterated, by means of a $2\times2$ matrix products. Indeed, introducing the
pair of numbers $(u_i,v_i)$ such that $A_i = \frac{u_i}{v_i}$, the 
recursion becomes
\[
\left[\begin{matrix}
u_{i+1}\\[0.4cm]
v_{i+1}
\end{matrix}\right] = 
\left[\begin{matrix}
\sqrt{\frac{\lambda_a^+(i+1)}{\lambda_a^-(i)}} 
& - \frac{J_a\n\bigl(\frac{i}{N}\bigr)}{\sqrt{\lambda_a^+(i+1)\lambda_a^-(i)}}
\\[0.4cm]
0 & \sqrt{\frac{\lambda_a^-(i)}{\lambda_a^+(i+1)}}
\end{matrix}\right] 
\left[\begin{matrix}
u_i\\[0.4cm]
v_i
\end{matrix}\right]
\egaldef
M_i \left[\begin{matrix}
u_i\\[0.4cm]
v_i
\end{matrix}\right] ,
\]    
where for convenience we divided everything by the common factor 
$\sqrt{\lambda_a^-(i)\lambda_a^+(i+1)}$.
Let us define the  matrices ($p$ being a positive integer)
\begin{eqnarray*}
G^0\bigl(\frac{i+p}{N},\frac{i}{N}\bigr) &\egaldef & \prod_{j=i}^{i+p}
\left[\begin{matrix}
\sqrt{\frac{\lambda_a^+(j+1)}{\lambda_a^-(j)}} 
& 0\\[0.4cm]
0
& \sqrt{\frac{\lambda_a^-(j)}{\lambda_a^+(j+1)}}
\end{matrix}\right] ,  \\[0.2cm]
G\bigl(\frac{i+p}{N},\frac{i}{N}\bigr) &\egaldef & \prod_{j=i}^{i+p}M_j, \\[0.2cm]
\Sigma\bigl(\frac{i}{N}\bigr) &\egaldef&
\left[\begin{matrix}
0 
& - \frac{J_a\n\bigl(\frac{i}{N}\bigr)}{\sqrt{\lambda_a^+(i+1)\lambda_a^-(i)}}
\\[0.4cm]
0 & 0
\end{matrix}\right],
\end{eqnarray*}
(explicit references to the species $(a)$ and the size $N$ is omitted here, to lighten 
the notations).
Because of the upper triangular structure of $\Sigma$, we may simply express
$G$ as
\[
\begin{split}
G\bigl(\frac{i+p}{N},\frac{i}{N}\bigr) = G^0\bigl(\frac{i+p}{N},\frac{i}{N}\bigr) 
+\sum_{j=0}^p G^0\bigl(\frac{i+p}{N},\frac{i+j+1}{N}\bigr)
\Sigma\bigl(\frac{i+j}{N}\bigr)
G^0\bigl(\frac{i+j-1}{N},\frac{i}{N}\bigr).
\end{split}
\]
To handle this equation in the continuous limit, we need an additional
transformation. Define
\[
L_i =  
\left[\begin{matrix}
\sqrt{\frac{\Gamma_a}{\lambda_a^+(i)}} & 0\\[0.4cm]
0 & \sqrt{\frac{\lambda_a^+(i)}{\Gamma_a}}
\end{matrix}\right] , \qquad 
R_i =  \left[\begin{matrix}
\sqrt{\frac{\lambda_a^-(i)}{\Gamma_a}} &0 \\[0.4cm]
0 & \sqrt{\frac{\lambda_a^-(i)}{\Gamma_a}}  
\end{matrix}\right], 
\]
together with
\begin{equation}\label{transform}
\begin{cases}
\DD\tilde G\bigl(\frac{i+p}{N},\frac{i}{N}\bigr) = L_{i+p+1}
 G\bigl(\frac{i+p}{N},\frac{i}{N}\bigr)  R_i\\[0.2cm]
\DD\tilde G^0\bigl(\frac{i+p}{N},\frac{i}{N}\bigr) = L_{i+p+1}
G^0\bigl(\frac{i+p}{N},\frac{i}{N}\bigr)  R_i
\end{cases}.
\end{equation}
Then $\tilde G$, $\tilde G^0$ and $\tilde\Sigma$ verify the same relation, 
\begin{equation}\label{Dyson}
\begin{split}
\tilde G\bigl(\frac{i+p}{N},\frac{i}{N}\bigr) = 
\tilde G^0\bigl(\frac{i+p}{N},\frac{i}{N}\bigr) 
+ \sum_{j=0}^p \tilde G^0\bigl(\frac{i+p}{N},\frac{i+j+1}{N}\bigr)
\tilde\Sigma\bigl(\frac{i+j}{N}\bigr)
\tilde G^0\bigl(\frac{i+j}{N},\frac{i+1}{N}\bigr),
\end{split}
\end{equation}
but
\[
\tilde\Sigma\bigl(\frac{i}{N}\bigr) =
\left[\begin{matrix}
0 & -\frac{\Gamma_a J_a\n(\frac{i}{N})}{\lambda_a^+(i+1)\lambda_a^-(i)}\\[0.4cm]
0 & 0
\end{matrix}\right].
\]
Noting that $A_{i+p+1}\Gamma_a/\lambda_a^+(i+p+1) = A_{i+p+1}$
and $A_i\Gamma_a/\lambda_a^-(i) = A_i$, the iteration between $i$
and $i+p$  gives
\begin{equation}\label{iterp}
A_{i+p+1} = \frac{\tilde G_{11} \bigl(\frac{i+p}{N},\frac{i}{N}\bigr)
A_i + \tilde G_{12}\bigl(\frac{i+p}{N},\frac{i}{N}\bigr)}
{\tilde G_{22}\bigl(\frac{i+p}{N},\frac{i}{N}\bigr)}.
\end{equation}
We can now take advantage of  the law of large numbers in equation (\ref{Dyson}). 
First of all, for $N$ large, and fixing $x=i/N$ and $y=p/N$, letting 
$\sigma = \left[\begin{matrix}1 & 0\\0&-1\end{matrix}\right]$,
we have,
\[
\tilde G^0\bigl(\frac{i+p}{N},\frac{i}{N}\bigr)
= \exp\Bigl(\frac{\sigma}{2}\sum_{j=i+1,k=2}^{i+p,n}
\log\frac{\lambda_{ak}}{\lambda_{ka}}X_j^k\Bigr)
= \exp\Bigl(\frac{\sigma}{2}\int_x^{x+y}du\sum_{k=2}^n\alpha^{ak}\rho_k(u)
+o(1) \Bigr),
\]
from the hydrodynamic hypothesis.
To proceed further, we have to distinguish between two situations.
\medskip
 
{\bf [The equidiffusion case]}

Recalling that $\Gamma_a$ is a free parameter which scales like $N$, 
it is convenient in the {\it equidiffusion} case to impose the limit
\[
\lim_{N\to\infty} \frac{\Gamma_a(N)}{N} = D.
\]
Then, expanding $\tilde\Sigma(i/N)$ with respect to $1/N$ yields
\[
\tilde\Sigma\bigl(\frac{i}{N}\bigr) = 
\left[\begin{matrix}
0 & -\frac{J_a\n(\frac{i}{N})}{ND}\\[0.4cm]
0 & 0
\end{matrix}\right]
+ {\cal O}\bigl(N^{-2}\bigr),
\]
and the limit
\[
{\cal G}(x+y,x) \egaldef \lim_{N\to\infty} \tilde G\bigl(\frac{i+p}{N},\frac{i}{N}\bigr)
\]
is provided by equation (\ref{Dyson}). Hence
\begin{equation}\label{Cdyson}
{\cal G}(x+y,x) = {\cal G}^0(x+y,x) +\int_x^{x+y}du\ 
{\cal G}^0(x+y,x+u)\ \Xi(x+u)\ 
{\cal G}^0(x+u,x),
\end{equation}
with 
\begin{equation}\label{freemat}
{\cal G}^0(y,x) =  
\exp\Bigl(\frac{\sigma}{2}\int_x^ydu\sum_{k=2}^n\alpha_{ak}\rho_k(u)
\Bigr),\qquad\text{and}\qquad \Xi(x) = \left[\begin{matrix}
0 & -\frac{J_a\n(x)}{D}\\[0.4cm]
0 & 0
\end{matrix}\right],
\end{equation}
still by virtue of the hydrodynamic hypothesis (\ref{hhydro}). 
Now it is possible to close the equations between densities and currents. Using again 
the hydrodynamic hypothesis with the fact that $\cal G$ is a smooth deterministic
operator, (\ref{iterp}) leads to,
\[
\rho_a(x+y) = \frac{{\cal G}_{11} (x+y,x)\rho_a(x)+ {\cal G}_{12}(x+y,x)}
{{\cal G}_{22}(x+y,x)}.
\]
By differentiating this last relation w.r.t.  $y$, 
altogether with (\ref{Cdyson}) and (\ref{freemat}), 
we obtain the final deterministic expression for the current
\begin{equation}\label{detercurrent}
J_a(x) = D\bigl(-\frac{\partial\rho_a}{\partial x}+
\sum_{k=2}^n\alpha_{ak}\rho_k\rho_a\Bigr),
\end{equation}
which, combined with the continuity equation 
\[
\frac{\partial \rho_a}{\partial t} +\frac{\partial J_a}{\partial x} = 0,
\]
leads again to a Burger's hydrodynamic equation. 

\medskip
[{\bf The hetero-diffusion case}] Here, the limit (\ref{Dyson}) is a bit 
more tricky. In fact, the expansion of $\tilde\Sigma$ involves correlations 
between currents 
and densities which already appear  in the leading terms, and we expect an 
effective diffusion constant of the form
\[
D_a(\rho) = D\exp\Bigl(\sum_{k=2}^n \beta^{ak}\rho_k\Bigr) ,
\]
with
\[
\begin{cases}
\DD D \egaldef \lim_{N\to\infty}\frac{1}{N^2}\exp\Bigl(\frac{1}{n-1}
\sum_{k=2}^n\log\lambda_{ak}(N)\Bigr) ,\\[0.2cm]
\DD\beta^{ak} \egaldef \lim_{N\to\infty}
\log\Bigl(\frac{\lambda_{ak}}{N^2 D}\Bigr).
\end{cases}
\]
We  pursue no further the study of this case, which presumably
could be handled with block-estimates techniques (see \cite{Sp}).

\subsubsection{Diffusion with reaction}
Here we treat the square-lattice model, a special case of 
(\ref{evnmod}),where reactions take place, in addition to 
diffusion. The procedure follows the lines of the preceding subsection.
Using the mapping (\ref{cpmapping}), the model is formulated in terms of
two coupled exclusion processes, and
the current equations corresponding to both species have the form
\begin{align*}
J_a\n\bigl(\frac{i}{N}) &= \lambda_a^+(i) \tau_i^a\bar\tau_{i+1}^a -
\lambda_a^-(i) \bar\tau_i^a\tau_{i+1}^a  , \\[0.2cm]
J_b\n\bigl(\frac{i}{N}) &= \lambda_b^+(i) \tau_i^b\bar\tau_{i+1}^b -
\lambda_b^-(i) \bar\tau_i^b\tau_{i+1}^b ,
\end{align*}
with the rates given by (\ref{taux}), and we restrict the present analysis to the
symmetric case (see relations (\ref{symmetric})). Reversing for example the 
equation for $J_a$ leads to the homographic relationship
\[
\tau_{i+1}^a = \frac{\lambda_a^+(i)\tau_i^a - J_a\n\bigl(\frac{i}{N}\bigr)}
{(\lambda_a^+(i)-\lambda_a^-(i))\tau_i^a+\lambda_a^-(i)},
\]
which again can be iterated by means of a $2\times 2$ matrix product, after defining 
$u_i^a$
and $v_i^a$ s.t. $\tau_i^a=u_i^a/v_i^a,$ $\forall i\in\{1\ldots N\}$.
Define
\[
\lambda(N) \egaldef \frac{\lambda^+(N)+\lambda^-(N)}{2},\qquad\mu(N)
\egaldef   \frac{\lambda^+(N)-\lambda^-(N)}{2},
\]
and
\[
\gamma(N) \egaldef \frac{\gamma^+(N)+\gamma^-(N)}{2}.
\]
Then the proper scalings for large $N$ are given by
\[
\lim_{N\to\infty}\frac{\lambda(N)}{N^2} = D , \qquad
\lim_{N\to\infty}\frac{\gamma(N)}{N^2} = \Gamma , \qquad
\lim_{N\to\infty}\frac{\mu(N)}{N} = \eta.  
\]
Letting now,
\[
\Sigma\bigl(\frac{i}{N}\bigr) =
\left[\begin{matrix}
0 & -\frac{J_a\n(\frac{i}{N})}
{\sqrt{\lambda_a^+(i)\lambda_a^-(i)}}\\[0.4cm]
\sqrt{\frac{\lambda_a^+(i)}{\lambda_a^-(i)}}
-\sqrt{\frac{\lambda_a^-(i)}{\lambda_a^i(i)}}& 0
\end{matrix}\right].
\]
$G$ cannot be given explicitly, it is instead solution of the following
combinatorial self-consistent equation
\begin{equation}\label{Dyson2}
G\bigl(\frac{i+p}{N},\frac{i}{N}\bigr) = 
G^0\bigl(\frac{i+p}{N},\frac{i}{N}\bigr)
+\sum_{j=0}^p G^0\bigl(\frac{i+p}{N},\frac{i+j}{N}\bigr)
\Sigma\bigl(\frac{i+j}{N}\bigr)\ G\bigl(\frac{i+j}{N},\frac{i+1}{N}\bigr).
\end{equation}
The iteration now reads,
\[
\left[\begin{matrix}
u_{i+p+1}\\[0.4cm]
v_{i+p+1}
\end{matrix}\right] = G\bigl(\frac{i+p}{N},\frac{i}{N}\bigr)
\left[\begin{matrix}
u_{i}\\[0.4cm]
v_{i}
\end{matrix}\right]
\]
For the same reason as before, the limit $\cal G$ of $G$ when $N\to\infty$ 
does satisfy 
\begin{equation}\label{dyson3}
{\cal G}(x+y,x) = {\cal G}^0(x+y,x) +\int_x^{x+y}du\ {\cal G}^0(x+y,x+u)\Sigma(x+u)
{\cal G}(x+u,x),
\end{equation}
with
\[
{\cal G}^0(y,x) =  
\exp\Bigl(\eta\sigma\int_x^y (2\rho_b(u)-1)du
\Bigr),
\]
by just applying the law of large numbers in  the formal expansion of $G$ with
respect to $\Sigma$. We leave aside the question concerning
existence and  analytic properties of a solution of (\ref{dyson3}).
We must again discriminate between two situations.

\medskip
{\bf [Case $\gamma=\lambda$]} 
\[
\Sigma(x) = 
\left[\begin{matrix}
\eta(2\rho_b-1) & -\frac{J_a(x)}{D}\\[0.4cm]
2\eta(2\rho_b-1) & \eta(1-2\rho_b)
\end{matrix}\right],
\]
which leads to the following differential system, 
\begin{align*}
\frac{\partial u^a}{\partial x} &= \eta(2\rho_b-1)u^a 
-\frac{1}{D}J_a(x)v^a  , \\[0.2cm]
\frac{\partial v^a}{\partial x} &= 2\eta(2\rho_b-1)u^a + \eta(1-2\rho_b)v^a , 
\end{align*}
after making use of the law of large numbers and the hydrodynamic hypothesis. 
Combining these last two equations to express 
$\rho_a'=(u_a'v_a-v_a'u_a)/v_a^2$,
leads to the relation
\[
J_a(x) = -D\Bigl(\frac{\partial\rho_a}{\partial x}+
2\eta\rho_a(1-\rho_a)(1-2\rho_b)\Bigr).
\]

{\bf [Case $\gamma\ne\lambda$]}

Like in the {\it hetero-diffusion} case of the last section, 
the effective diffusion constant $D_a(\rho)$ involves correlations between 
$\tau_i^b$ and $\tau_{i+1}^b$ and $J_a(i/N)$ in the leading
order term, and we expect a behavior of the form \cite{FaFu}
\[
D_a(\rho_b) = D\exp\Bigl[2\rho_b(1-\rho_b)\log\frac{\gamma}{\lambda}\Bigr],
\]
as a result of a multiplicative process. This could  be obtained
through renormalization techniques applied directly to equation (\ref{Dyson2}). 

To conclude  this section, we see that, for $\gamma=\lambda$, 
the  differential system expressing, at steady state, the deterministic  limit of
the  square lattice  model  with periodic  boundary conditions 
finally reads, setting $\nu_{a,b} = 2\rho_{a,b}-1$,
\begin{equation}\label{nldiff}
\begin{cases}
\DD\frac{\partial\nu_a}{\partial  x}  =  \eta(1-\nu_a^2)\nu_b\  +\
v\nu_a+\varphi^a,  \\[0.2cm]   \DD\frac{\partial\nu_b}{\partial  x}  =  -
\eta(1-\nu_b^2)\nu_a\ +\ v\nu_b+\varphi^b,
\end{cases}
\end{equation}
where $v$ is a possibly finite drift velocity and  $\varphi^a=\varphi(\bar\nu_a,\bar\nu_b)$ 
and $\varphi^b(\bar\nu_a,\bar\nu_b)$  are two constant currents in the translating frame.
These currents have to be determined in a self-consistent manner, after fixing  
the average densities $\bar\nu_a$ and $\bar\nu_b$ and the periodic boundary 
conditions.
For $v=0$, the  system (\ref{nldiff}) is Hamiltonian with 
\begin{equation}
H = \frac{\eta}{2}\bigl[\nu_a^2\nu_b^2 - \nu_a^2-\nu_b^2\bigr]+
\varphi_b\nu_a-\varphi_a\nu_b.
\end{equation}
Indeed, it is easy to observe that (\ref{nldiff}) can be rewritten as
\[
\frac{\partial\nu_a}{\partial x} = -\frac{\partial H}{\partial\nu_b}, \qquad
\frac{\partial\nu_b}{\partial x} =  \frac{\partial H}{\partial\nu_a}. 
\]
The  degenerate fixed point $\nu_{a,b}(x)=\bar\nu_{a,b}$  is always a trivial 
solution and corresponds to the relations
\[
\varphi_a = \eta(\bar\nu_a^2 - 1)\bar\nu_b, \qquad \varphi_b = 
\eta(1-\bar\nu_b^2)\bar\nu_a.
\]

\subsection{Microscopic currents}
\subsubsection{Particle currents}
An important feature of our particle systems is that the number of 
particles is locally conserved. 
This property is reflected as $N\to\infty$ by a continuity equation, 
which relates local variations of  particle density to inhomogeneous 
currents. In a discretized framework, conservation of 
particles is expressed according to the following
\begin{prop}\label{condprodform}
Let $\{J_i^k(t,\epsilon)\}\ i=1,\ldots, N$ be stochastic variables 
corresponding to the fluxes of particles of type $k\in\{1,\ldots, n\}$ 
between site $i$ and $i+1$, such that
\[
J_i^k(t,\epsilon) \egaldef \frac{1}{\epsilon}\sum_{l\ne k} 
\Bigl( X_i^k(t) X_{i+1}^l(t)X_i^l(t+\epsilon) X_{i+1}^k(t+\epsilon)-
X_i^l(t) X_{i+1}^k(t)X_i^k(t+\epsilon) X_{i+1}^l(t+\epsilon)\Bigr)
\]
with $\epsilon>0$. By definition 
$ J_i^{k}(t,\epsilon)$ are  ternary variables
in $ \{-\frac{1}{\epsilon},0,+\frac{1}{\epsilon}\}$. 
The following identity, equivalent to particle conservation, 
\begin{equation}\label{Ward}
\lim_{\epsilon\to 0} \frac{X_i^k(t+\epsilon)-
X_i^k(t)}{\epsilon} +J_{i+1}^k(t,\epsilon) - J_i^k(t,\epsilon) = 0  \qquad a.s.,
\end{equation}
holds for all  $i\in\{1,\ldots, N\}$, $ \forall t\in{\mathbb R}^+$. 
In addition, letting $\eta\n(t)$ denote
the sequence $\{X_i^k(t)\},\ i=1,\ldots, N; k=1,\ldots ,n\}$, then the variables
$\{J_i^{k}(t,\epsilon)\},\ i=1,\ldots, N; k =1,\ldots, n\}$, 
have a joint conditional  Laplace transform given by
\begin{align}\label{caract}
h_{t,\epsilon}\n(\phi) &\egaldef 
\EE_t\left(\exp\Bigl(\frac{1}{N}\sum_{k<l\atop i=1}^{n,N}
\phi_k(\frac{i}{N})\epsilon 
J_i^k(t,\epsilon)\Bigr)\Big\vert \eta(t)\right) =  \nonumber \\[0.2cm]
&\EE_t\biggl[\exp\Bigl(\epsilon\sum_{k\ne l\atop i=1}^{n,N} 
\lambda_{kl}X_i^kX_{i+1}^l\bigl(e^{\frac{1}{N}\psi_{kl}(\frac{i}{N})}-1\bigr)
+\lambda_{lk}X_i^lX_{i+1}^k\bigl(e^{-\frac{1}{N}\psi_{kl}(\frac{i}{N})} -1\bigr)   
\Bigr)\biggr]+\  o(\epsilon),
\end{align}
where $\phi_k,\ k =1,\ldots, n$ is a set   of $\mathbf{C^{\infty}}$ bounded functions, and
$\psi_{kl} = \phi_k-\phi_l$.
\end{prop}
\begin{proof}
The points are mere consequences of the Markovian feature of the process and of its 
generator. In particular, (\ref{Ward})  results  
from the fact that, almost surely, at most  one jump takes place in the 
time-interval $\epsilon$, when
$\epsilon\to 0$, since all events are due to independent Poisson processes. In addition,
on the time interval $[t,t+\epsilon]$, the occurrence of a particle exchange between 
sites $i$ and $i+1$, corresponding to $\epsilon J_i^k(t,\epsilon) = 1$ 
is only conditioned by the presence of a pair $(k,l)$  at $(i,i+1)$, with a 
transition rate given by $\lambda_{kl}X_i^kX_{i+1}^l$. Therefore 
\[
h_{t,\epsilon}\n(\phi) = {\mathbb E}_t
\biggl(\prod_{k\ne l\atop i=1}^{n,N}\Bigl[1+\epsilon\lambda_{kl}X_i^kX_{i+1}^l
\bigl(e^{\frac{1}{N}\psi_{kl}(\frac{i}{N})}-1\bigr)\Bigr]\biggr), 
\]
which, after a first order expansion with respect to $\epsilon$, leads to (\ref{caract}) . 
\end{proof}

\subsubsection{An iterative numerical scheme}
Given a sample path $\eta\n(t)$ at time $t$, 
we may generate a current sequence $\{J_i^k(t,\epsilon)\}$ 
according to the local product form encountered earlier. In turn, once the set  
$\{J_i^k(t,\epsilon)\}$ is known, the sequence $\eta(t+\epsilon)$ is almost 
surely determined, as $\epsilon\to 0$, 
by the identity (\ref{Ward}), expressing  conservation law
of particles. We therefore have at hand an explicit stochastic
numerical scheme to generate the sequence $\eta(t)$ step by step. 
\begin{prop}
For any $\epsilon>0$, $N\in {\mathbb N}$, the iterative scheme given by
\[
Q_{n+1}(\eta) = \sum_{\eta'} P_\epsilon(\eta \vert \eta') Q_n(\eta),
\]
where $P_\epsilon(\eta \vert \eta')$ is defined according to (\ref{Ward}) and
(\ref{caract}), converges  when $\epsilon\to 0$ to the original probability 
measure $P_{t=n\epsilon}(\eta)$ corresponds to the original process.
\end{prop}      
\begin{proof}
There is only one thing to show: $\forall T>0$, the probability $p_\epsilon$ that
$\exists\,t\in[0,T]$, such that two adjacent transitions occur within
the same time-interval $[t,t+\epsilon]$, tends to $0$ when $\epsilon\to 0$. 
This is warranted by the fact that the total number
of transitions for $t<T$ is almost certainly finite. Indeed, 
we have
\[
p_\epsilon\le 1- \bigl(1-(\max_{kl}\lambda_{kl})^2
\epsilon^2\bigr)^{\frac{NT}{\epsilon}} 
\ra_{\epsilon\to 0} 0.
\]
For the hydrodynamic limit the rates $\lambda_{kl}$ 
scale like $N^2$ for large $N$. Thus, it will be  convenient
to take a single limit  $\epsilon \egaldef \epsilon(N)\to 0 $ as $N \to \infty$,
since the condition for the scheme to be meaningful  writes
\[
N\epsilon(N) (\max_{kl}\lambda_{kl})^2 = o(1),
\]
so that we get a scaling of $\epsilon(N) = o(N^{-5})$ to meet our 
needs. This will allow us, in the sequel, to make use of the approximation
\[
\sum_{i=1}^N \alpha_i^k\Bigl(X_i^k(t+\epsilon)-X_i^k(t)
-\sum_l\bigl(J_{i-1}^k-J_i^k\bigr)\epsilon\Bigr) = 
o(\epsilon),
\]
for any set of bounded complex numbers $\{\alpha_i^k\}$.
\end{proof}

\subsubsection{Central limit theorem for currents}\label{fluctu}
We are  in position to exploit the conditional product form (\ref{caract})
to perform a mapping, in the spirit of Lemma 4.1 of \cite{FaFu} ,
allowing to obtain a dynamical description of the system, in terms of some external 
free random process. To this end we  assume, as a basic point,  the hydrodynamic limit holds and we rest on the following lemma. 
\begin{lem} 
Suppose  the existence of a set of density functions $\rho_k$,
such that
\[
{\mathbb E}\Bigl[\exp\bigl(\frac{1}{N}\sum_{k=1\atop i=1}^{n,N}
X_i^k\phi_k\bigl(\frac{i}{N}\bigr)\bigr)\Bigr]
= \exp\Bigl(\sum_{k=1\atop i=1}^{n,N}\log\Bigl[1+\rho_k\bigl(\frac{i}{N}\bigr)
\bigl(e^{\phi_k\bigl(\frac{i}{N}\bigr)}-1\bigr)\Bigr]+ o(N^{-2})\Bigr),
\]
for any given bounded complex function $\phi_k$, and let 
 $\DD \phi  = \sup_{k\in\{1\ldots n\}\atop x\in[0,1]}(\phi_k(x))$. Then, 
\[
{\mathbb E}\Bigl[\exp\bigl(\frac{1}{N}\sum_{k<l\atop i=1}^N
\phi_k(\frac{i}{N})\phi_l(\frac{i}{N})X_i^kX_i^l\bigr)\Bigr] =
\exp\Bigl(\frac{1}{N}\sum_{k<l\atop i=1}^N \phi_k(\frac{i}{N})\phi_l(\frac{i}{N}) 
\rho_k\bigl(\frac{i}{N}\bigr)\rho_l\bigl(\frac{i}{N}\bigr)
\ +\ o\bigl(\frac{\phi}{N}\bigr)\Bigr).
\]
\end{lem}
From this we deduce the following identity,
\begin{align}
h_{t,\epsilon}\n(\phi) = \exp\Bigl(
\epsilon\sum_{k<l\atop i=1}^{n,N} 
\lambda_{kl}\rho^k\bigl(\frac{i}{N}\bigr)
\rho^l\bigl(\frac{i+1}{N}\bigr)\bigl(e^{\frac{1}{N}
\psi_{kl}(\frac{i}{N})}-1\bigr) 
+\lambda_{lk}\rho^l\bigl(\frac{i}{N}\bigr)
\rho^k\bigl(\frac{i+1}{N}\bigr)\bigl(e^{-\frac{1}{N}\psi_{kl}(\frac{i}{N})}-1\bigr)   
\Bigr)+o(\epsilon)\Bigg),
\end{align}
which leads to recover (in our specific context) a formulation of 
the general result of \cite{BeLa} concerning fluctuation laws of currents 
for diffusive systems. 

Keeping up to quadratic terms w.r.t. to functions $\phi$'s its
argument, $h_{t,\epsilon}\n(\phi)$ reads,
\begin{equation}\label{heps}
h_{t,\epsilon}\n(\phi) = \exp\Bigl(\epsilon\sum_{k=1\atop i=1}^{n,N}
\phi_k(\frac{i}{N}){\cal J}^k(\rho(\frac{i}{N})) +
\frac{D\epsilon}{N^2}\sum_{k,l=1}^n 
\phi_k\bigl(\frac{i}{N}\bigr) Q_{kl}\bigl(\frac{i}{N}\bigl)\phi_l\bigl(\frac{i}{N}\bigr)
+o(\frac{\phi^2}{N^2})\Bigr)
\end{equation}
where ${\cal J}^k$ are  deterministic currents expressed, in terms of densities, by 
\[
{\cal J}^k(\{\rho_l,l=1\ldots n\}) \egaldef -D\Bigl(\frac{\partial\rho_k}{\partial x} 
+\sum_{l\ne k}\alpha_{kl}\rho_k\rho_l\Bigr),
\]
and $Q$ is a 
$n\times n$ symmetric matrix 
\[
\begin{cases}
Q_{ij} = -\rho_i\rho_j,\qquad i\ne j, \\[0.2cm]
Q_{ii} = \rho_i(1-\rho_i).
\end{cases}
\]
$Q$ is of rank $n-1$, because due to the exclusion constraint, currents are not 
independants,
\[
\sum_{k=1}^n J_i^k(t,\epsilon) = 0, \qquad\forall i\in\{1,\ldots,N\}.
\]
Let $M$ the reduced matrix obtained from $Q$ by deleting last row and last 
column.
Its  determinant is $\prod_{k=1}^n\rho_k$, so that it 
$M$ invertible if none of the $\rho_k$ vanishes, with
\begin{equation}\label{minverse}
\begin{cases}
\DD M^{-1}_{ij} = \frac{1}{\rho_i}+\frac{1}{\rho_n},\qquad i\ne j , \\[0.2cm]
\DD M^{-1}_{ii} = \frac{1}{\rho_n},
\end{cases}
\end{equation}
after having taken into account the exclusion condition $\sum_{k=1}^n \rho_k=1$.
Since every line $k$ or  column $k$ sums to $\rho_k\rho_n >0$, all the eigenvalues 
are strictly positive, and hence $M(\rho)$  ows a real square-root 
matrix $M^{\frac{1}{2}}(\rho)$.
\begin{prop}
Let $\phi_k, k=1,\ldots, n-1$ denote a set  of $C^\infty$ bounded  functions 
of the real variable $x\in [0,1]$,  $\{w_i^k,k=1,\ldots, n-1\}$ 
a set of independent identically distributed 
Bernoulli  random variables with parameters $1/2$, taking at time $t$ 
values in $\{-1/2,1/2\}$.
Then there exists a probability space, such that
\begin{equation}\label{currentmapping}
\frac{1}{N} \sum_{k=1\atop i=1}^{n,N}\ \phi_k(\frac{i}{N}) J_i^k\epsilon  = 
\frac{1}{N} \sum_{k=1\atop i=1}^{n-1,N}\ \psi_{kn}(\frac{i}{N})\Bigl
[{\cal J}^k(\rho(\frac{i}{N}))\epsilon +\sqrt{2D\epsilon} \sum_{l=1}^{n-1} 
M_{kl}^\frac{1}{2}\Bigl(\rho\bigl(\frac{i}{N}\bigr)\Bigr)\ w_i^l\Bigr]
+{\cal O}(N^{-2}),\ a.s., 
\end{equation}
\end{prop}
The lines of arguments bare some features in common with the ones proposed in 
\cite{FaFu} (to study  fluctuations at steady state).
Recall, by law of large numbers,  that correlations are negligible
and do not affect the expression of the  deterministic currents (\ref{detercurrent}).
This justifies the mapping (\ref{currentmapping}). On the other hand, the calculation of 
coefficents $M_{ij}^{\frac{1}{2}}$ is done  by comparing $h_{t,\epsilon}\n$ in (\ref{heps}) 
with
\[
{\mathbb E}\Bigl[\exp\bigl(\frac{1}{N}\sum_{k=1\atop i=1}^{n-1,N}
\psi_{kn}\bigl(\frac{i}{N}\bigr)  \sqrt{2D\epsilon} 
M^\frac{1}{2}_{kl}(\frac{i}{N})w_i^l\bigr)\Bigr]
= 
\exp\Bigl(\frac{D\epsilon}{N^2}\sum_{kl}^n 
\phi_k\bigl(\frac{i}{N}\bigr) Q_{kl}\bigl(\frac{i}{N}\bigl)\phi_l\bigl(\frac{i}{N}\bigr)
+o(\epsilon)\Bigr),
\]
because  $M^\frac{1}{2}$ is symmetric and 
\[
\sum_{kl}^{n-1} \psi_{kn}\bigl(\frac{i}{N}\bigr)M_{kl}\psi_{ln}\bigl(\frac{i}{N}\bigr)
= \sum_{kl}^n 
\phi_k\bigl(\frac{i}{N}\bigr) Q_{kl}\bigl(\frac{i}{N}\bigl)\phi_l.
\]
Setting, for $k=1,\ldots,n-1$,
\[
Y_k\n(x,t) \egaldef \frac{1}{\sqrt N}\sum_{i=1}^{[xN]} w_i^k,
\]
the corresponding space time white noise processes
\[
W^k(x,t) = \lim_{N\to\infty} \frac{d Y_k\n}{dx}(x,t),
\]
describe current fluctuations in the continuous limit. 

\subsection{Macroscopic fluctuations}
Two main quantities with be explored in this section:  the 
Lagrangian and the large deviation functional.

\subsubsection{The Lagrangian}
The preceding section provides us with all coefficients required to achieve an informal 
derivation of  the Lagrangian \cite{BeLa}  describing the current  fluctuations. Given the 
empirical measure
\[
\rho_k\n(x,t) \egaldef \frac{1}{N}\sum_{i=1}^n X_i^k(t)
\delta\bigl(x-\frac{i}{N}\bigr), 
\]
and assuming the system admits a hydrodynamical  description in terms
of a density field $\rho_k(x,t)$, the statement in \cite{BeLa} says that there is  
a large deviation principle for the stationary measure. In other words, 
the probability that the measure $\rho_k\n$ deviates from the hydrodynamic 
density profile  $\rho_k$ is exponentially small and given by
\[
P\bigl\{\rho\n(t)\simeq\hat\rho(t),t\in[t_1,t_2]\bigr\}
\simeq e^{-NI_{[t_1,t_2]}(\hat\rho)},
\]  
where
\[
I_{[t_1,t_2]}(\hat\rho) = \int_{t_1}^{t_2}{\cal L}(\hat\rho(t),\partial_t\hat\rho(t)) dt .
\]
Here the deviation from hydrodynamic solutions is due to current fluctuations.\\
Writing  $\bigtriangledown^{-1}\egaldef\int_0^x$, the quantity 
$\bigtriangledown^{-1}\frac{\partial\hat\rho_k\n}{\partial t} + {\cal J}^k(\hat\rho)$,
represents the fluctuations of the current $J^k$. Reversing the relationship between 
current 
fluctuations and  white noise process leads formally to 
\begin{equation}\label{noise}
W^l(x,t) \simeq  \sqrt\frac{\epsilon}{2DN}
\sum_{k=1}^{n-1} M^{-\frac{1}{2}}_{lk}\Bigl(
\bigtriangledown^{-1}\frac{\partial\hat\rho_k}{\partial t} 
+ {\cal J}_k(\hat\rho)\Bigr) , \quad l=1,\ldots, n-1.
\end{equation}
Then, replacing (\ref{noise}) in 
the joint distribution  of $\{W^k(x,t); x\in[0,1],k=1,\ldots,n-1\}$ , we obtain
\begin{align*}
{\cal L}(\hat\rho(t),\partial_t\hat\rho(t))dt
&= \frac{1}{2}\int_0^1 dxdt \sum_{k=1}^{n-1}  \Bigl(W^k(x,t)\Bigr)^2 \\[0.2cm]
&=  \frac{1}{4D}\int_0^1 dxdt \sum_{k=1}^{n-1} \Bigl(\sum_{l=1}^{n-1} M^{-\frac{1}{2}}_
{lk}
\bigtriangledown^{-1}\frac{\partial\hat\rho_k}{\partial t} 
+ {\cal J}_k(\hat\rho)\Bigr)^2,
\end{align*}
where $\epsilon$ has been  identified with $dt$ and $dx$ with  $1/N$. 
Then, the symmetry of $M^{-\frac{1}{2}}$, the form  (\ref{minverse}) of $M^{-1}$
and the exclusion constraint
\[
\sum_{k=0}^n \bigtriangledown^{-1}\frac{\partial\hat\rho_k}{\partial t} 
+ {\cal J}_k(\hat\rho) = 0,
\]
lead to the final compact form
\[
\mathcal{L}(\hat\rho,\partial_t\hat\rho)
=\frac{1}{4D}\int_0^1 dx \sum_{k=1}^n \frac{ 
\Bigr(\bigtriangledown^{-1}\frac{\partial\hat\rho_k}{\partial t} 
+ {\cal J}_k(\hat\rho)\Bigl)^2}{\hat\rho_k} .
\]

\subsubsection{Hamilton-Jacobi equation and large deviation functional}
Here we proceed as in \cite{BeLa}.
Let $\pi_k$, the conjugate variable of $\rho_k$,
\[
\pi_k(x,t) \egaldef \frac{\partial\mathcal{L}(\rho,\partial_t\rho)}
{\partial\partial_t\rho_k(x,t)}.
\]
The Hamiltonian is then given by 
\[
{\cal H}(\{\rho_k,\pi_k\}) 
\egaldef \int_0^1 dx\sum_{k=1}^n\pi_k(x,t)\partial_t\rho_k(x,t) - {\cal L} .
\]
Algebraic manipulations lead to the expression 
\[
{\cal H}(\{\rho_k,\pi_k\}) = \int_0^1 dx 
\Bigl[\partial_x\pi_k {\cal J}_k(\rho)  +D\rho_k
\Bigl(\partial_x\pi_k\Bigr)^2 \Bigr].
\]
Then the large deviation functional $\mathcal{F}$, satisfying 
\[
P(\rho\n\simeq\rho) \simeq e^{-N \mathcal{F}(\rho)}, 
\]
might be derived as in \cite{BeLa}, from the following regular variational principle
\[
\mathcal{F}(\rho) = \inf_{\hat\rho}I_{[-\infty,0]}(\hat\rho),
\]
where the minimum is taken over all trajectories $\hat\rho$ connecting 
the stationary deterministic equilibrium profiles $\bar\rho_k$ to $\rho$. 
This means that $\mathcal{F}$  and  the action functional $I$ must satisfy 
the related Hamilton-Jacobi equation 
\[
{\cal H}\Bigl(\{\rho_k,\frac{\partial{\cal F}}{\partial\rho_k}\}\Bigr) = 0.
\]
In addition, one can check the relation
\[
{\cal F} = {\cal U}  - {\cal S},
\]
where  
\[
\begin{cases}
\DD {\cal U} = \int_0^1dx\int_0^x\sum_{k\ne l}
\alpha_{kl}\rho_k(x)\rho_l(y)dy,\\[0.3cm]
\DD {\cal S} = -\int_0^1 dx\sum_{k=1}^n\rho_k\log\rho_k,
\end{cases}
\]
a form already encountered  in the reversible case, see equation (\ref{freen}).
Indeed, when the process is reversible, $\cal U$ is  translation invariant 
(i.e. independent of the initial integration point, here set to zero), and so   
\[
\partial_x\frac{\partial {\cal F}}{\partial\rho_k(x)} = -\frac{{\cal J}_k}{D\rho_k}.
\]
This approach could be used to analyse the non-reversible case.

\section{Concluding Remarks}
In this report we strove to put forward some techniques,  and to extend methods to tackle the problem of  mapping discrete model to continuous equations. Even in the context of a very specific model, namely  stochastic distortions of discrete curves,  some difficult questions remain.
\begin{itemize}
\item[$\bullet$]  The determination of the invariant measure
in the general case, at the discrete level, which would generalize the totally asymmetric
case \cite{FeMa,MaMaRa}.  
\item[$\bullet$] The analysis of  Hamilton-Jacobi equations to  obtain a kind of continuous counterpart of the invariant measures, namely large deviation functionals. 
\end{itemize}

With regard to  hydrodynamic limits, there is a puzzling issue, namely when  
particle-species diffuse at various speeds, in what we called the  \emph{heterodiffusive} case. For many one-dimensional models, it is well known that a single slow particle
may considerably modify the macroscopic behavior of the system (see e.g. \cite{Ma}). 
 For the time being, our approach is restricted to  diffusive one-dimensional systems. 
Yet, other scalings (like Euler), as well as processes in higher dimension,  are definitely worth being studied. in particular, it could be interesting to deal with more realistic exclusion processes, for instance those encountered in the field of  traffic modelling. Besides, the analysis of irreversible invariant states in terms of cycles in a state-graph might well be extended to tackle  \textsc{asep}  on closed networks.

\nocite* 
\bibliography{refer} 

\begin{thebibliography}{10}

\bibitem{ArHeRi}
{\sc P.~Arndt, T.~Heinzel, and V.~Rittenberg}, {\em Stochastic models on a ring
  and quadratic algebras. the three-species diffusion problem}, J. Phys. A:
  Math. Gen., 31 (1998), pp.~833--843.

\bibitem{Berge}
{\sc C.~Berge}, {\em Th\'eorie des Graphes et ses Applications}, vol.~II of
  Collection Universitaire des Mathématiques, Dunod, 2ème~ed., 1967.

\bibitem{BeLa}
{\sc L.~Bertini, A.~De~Sole, D.~Gabrielli, G.~Jona~Lasinio, and C.~Landim},
  {\em Current fluctuations in stochastic lattice gases}, Phys. Rev. Lett., 94
  (2005), p.~030601.

\bibitem{Bil}
{\sc P.~Billingsley}, {\em Convergence of Probability Measures}, Wiley Series
  in Probability and Statistics, John Wiley \& Sons Inc., second~ed., 1999.

\bibitem{Bu}
{\sc J.~Burgers}, {\em A mathematical model illustrating the theory of
  turbulences}, Adv. Appl. Mech., 1 (1948), pp.~171--199.

\bibitem{ClDeEv}
{\sc M.~Clincy, B.~Derrida, and M.~R. Evans}, {\em Phase transition in the
  {ABC} model}, Phys. Rev. E, 67 (2003), pp.~6115--6133.

\bibitem{MaPr}
{\sc A.~De~Masi and E.~Presutti}, {\em Mathematical Methods for Hydrodynamic
  Limits}, vol.~1501 of Lecture Notes in Mathematics, Springer-Verlag, 1991.

\bibitem{DeEnLe}
{\sc B.~Derrida, C.~Enaud, and J.~L. Lebowitz}, {\em The asymmetric exclusion
  process and brownian excursions}, J. Stat. Phys., 115 (2004), pp.~365--382.

\bibitem{DeEvHaPa}
{\sc B.~Derrida, M.~R. Evans, V.~Hakim, and V.~Pasquier}, {\em Exact solution
  for 1d asymmetric exclusion model using a matrix formulation}, J. Phys. A:
  Math. Gen., 26 (1993), pp.~1493--1517.

\bibitem{DeMa}
{\sc B.~Derrida and K.~Mallick}, {\em Exact diffusion constant for the
  one-dimensional partially asymmetric exclusion model}, J. Phys. A: Math.
  Gen., 30 (1997), pp.~1031--1046.

\bibitem{EtKu}
{\sc S.~Ethier and T.~Kurtz}, {\em Markov Processes, Characterization and
  Convergence}, John Wiley {\&} Sons, 1986.

\bibitem{EvFoGoMu}
{\sc M.~R. Evans, D.~P. Foster, C.~Godrèche, and D.~Mukamel}, {\em Spontaneous
  symmetry breaking in a one dimensional driven diffusive system}, Phys. Rev.
  Lett., 74 (1995), pp.~208--211.

\bibitem{EvKaKoMu}
{\sc M.~R. Evans, Y.~Kafri, M.~Koduvely, and D.~Mukamel}, {\em Phase
  {S}eparation and {C}oarsening in one-{D}imensional {D}riven {D}iffusive
  {S}ystems}, Phys. Rev. E., 58 (1998), p.~2764.

\bibitem{FaFu}
{\sc G.~Fayolle and C.~Furtlehner}, {\em Dynamical {W}indings of {R}andom
  {W}alks and {E}xclusion {M}odels. {P}art {I}: Thermodynamic limit in $\mathbb
  {Z}^2$}, Journal of Statistical Physics, 114 (2004), pp.~229--260.

\bibitem{FaFu2}
\leavevmode\vrule height 2pt depth -1.6pt width 23pt, {\em Stochastic
  deformations of sample paths of random walks and exclusion models}, in
  Mathematics and computer science. III, Trends Math., Birkh\"auser, Basel,
  2004, pp.~415--428.

\bibitem{FaFu3}
\leavevmode\vrule height 2pt depth -1.6pt width 23pt, {\em Stochastic dynamics
  of discrete curves and exclusion processes. part 1: Hydrodynamic limit of the
  \textsc{asep} system}, Rapport de Recherche 5793, Inria, 2005.

\bibitem{FeMa}
{\sc P.~Ferrari and J.~Martin}, {\em Stationary distribution of multi-type
  totally asymmetric exclusion processes}.
\newblock math.PR/0501291.

\bibitem{GoLu}
{\sc C.~Godr\`eche and J.~Luck}, {\em Nonequilibrium dynamics of urns models},
  J. Phys. Cond. Matter, 14 (2002), p.~1601.

\bibitem{Ka}
{\sc O.~Kallenberg}, {\em Foundations of Modern Probability}, Springer, second
  edition~ed., 2001.

\bibitem{KPZ}
{\sc M.~Kardar, G.~Parisi, and Y.~Zhang}, {\em Dynamic scaling of growing
  interfaces}, Phys. Rev. Lett., 56 (1986), pp.~889--892.

\bibitem{Kel}
{\sc F.~P. Kelly}, {\em Reversibility and stochastic networks}, John Wiley \&
  Sons Ltd., 1979.
\newblock Wiley Series in Probability and Mathematical Statistics.

\bibitem{KiLa}
{\sc C.~Kipnis and C.~Landim}, {\em Scaling limits of Interacting Particles
  Systems}, Springer-Verlag, 1999.

\bibitem{LaBaRa}
{\sc R.~Lahiri, M.~Barma, and S.~Ramaswamy}, {\em Strong phase separation in a
  model of sedimenting lattices}, Phys. Rev. E, 61 (2000), pp.~1648--1658.

\bibitem{Li}
{\sc T.~M. Liggett}, {\em Stochastic Interacting Systems: Contact, Voter and
  Exclusion Processes}, vol.~324 of Grundlehren der mathematischen
  {W}issenschaften, Springer, 1999.

\bibitem{Ma}
{\sc K.~Mallick}, {\em Shocks in the asymmetry exclusion model with an
  impurity}, J. Phys. A: Math. Gen., 29 (1996), pp.~5375--5386.

\bibitem{MaMaRa}
{\sc K.~Mallick, S.~Mallick, and N.~Rajewsky}, {\em Exact solution of an
  exclusion process with three classes of particles and vacancies}, J. Phys. A:
  Math. Gen., 32 (1999), pp.~8399--8410.

\bibitem{MUR}
{\sc J.~Murray}, {\em Mathematical Biology}, vol.~19 of Biomathematics,
  Springer-Verlag, second~ed., 1993.

\bibitem{RUD}
{\sc W.~Rudin}, {\em Functional Analysis}, International Series in Pure and
  Applied Mathematics, McGraw-Hill, second~ed., 1991.

\bibitem{Pi}
{\sc P.~S.}
\newblock private communication.

\bibitem{Sp}
{\sc H.~Spohn}, {\em Large Scale Dynamics of Interacting Particles}, Springer,
  1991.

\end{thebibliography}
\bibliographystyle{siam}

\end{document}